\documentclass[preprintnumbers,superscriptaddress,endnote,nofootinbib,aps,prd,floatfix,11pt]{revtex4}

\usepackage[utf8]{inputenc}

\usepackage{footmisc,enumerate}
\usepackage{multirow}
\usepackage{subfigure}
\usepackage{amsmath,amssymb}
\usepackage{graphicx}
\usepackage{placeins}
\usepackage{bbm}
\usepackage{xspace,slashed}
\usepackage{hyperref}
\hypersetup{colorlinks=true, citecolor=blue, urlcolor=blue, linkcolor=blue}
\usepackage[normalem]{ulem}

\usepackage{amsmath}
\usepackage{amsfonts}
\usepackage{amssymb}
\usepackage{graphicx}
\usepackage{cancel}
\usepackage{xcolor}
\usepackage{multirow}
\usepackage{pifont}

\usepackage{threeparttable}
\usepackage{slashed}
\usepackage{bbold}
\newcommand{\levifo}{\epsilon_{\mu\nu\rho\sigma}}

\usepackage[autostyle]{csquotes}

\newcommand{\tx}[1]{{#1}}

\newcommand{\D}{D\mkern-11.5mu/}

\begin{document}

\newcommand{\og}{\ensuremath{\tilde{O}_g}\xspace}
\newcommand{\ot}{\ensuremath{\tilde{O}_t}\xspace}

\providecommand{\abs}[1]{\lvert#1\rvert}

\newcommand{\Znunujets}{(Z\to{\nu\bar{\nu}})+\text{jets}}
\newcommand{\Welnujets}{(W\to{\ell\nu})+\text{jets}}
\newcommand{\Znunujet}{(Z\to{\nu\bar{\nu}})+\text{jet}}
\newcommand{\Welnujet}{(W\to{\ell\nu})+\text{jet}} 

\newcommand{\cw}{\ensuremath{C_{\widetilde{W}}\xspace}}
\newcommand{\chwb}{\ensuremath{C_{H\widetilde{W}B}}\xspace}
\newcommand{\chb}{\ensuremath{C_{H\widetilde{B}}}\xspace}
\newcommand{\chw}{\ensuremath{C_{H\widetilde{W}}}\xspace}
\newcommand{\cwlam}{  \ensuremath{\frac{C_{\widetilde{W}}}{\Lambda^2}\xspace}}
\newcommand{\chwblam}{\ensuremath{\frac{C_{H\widetilde{W}B}}{\Lambda^2}}\xspace}
\newcommand{\chblam}{ \ensuremath{\frac{C_{H\widetilde{B}}}{\Lambda^2}}\xspace}
\newcommand{\chwlam}{ \ensuremath{\frac{C_{H\widetilde{W}}}{\Lambda^2}}\xspace}
\newcommand{\Reylyr}{\ensuremath{\text{Re}(y_{\chi_L} y^*_{\chi_R})}\xspace}
\newcommand{\Imylyr}{\ensuremath{\text{Im}(y_{\chi_L} y^*_{\chi_R})}\xspace}

\setlength{\footnotesep}{0.5cm}

\title{Landscaping CP-violating BSM scenarios}

\begin{abstract}
We consider a wide range of UV scenarios with the aim of informing searches for CP violation at the TeV scale using
effective field theory techniques. We demonstrate that broad theoretical assumptions about the nature of UV dynamics
responsible for CP violation map out a small subset of relevant operators at the TeV scale. Concretely, this will allow us to
reduce the number of free parameters that need to be considered in experimental investigations, thus enhancing analyses' sensitivities. In parallel, reflecting the UV dynamics' Wilson coefficient hierarchy will enable a streamlined theoretical interpretation of such analyses in the future. We demonstrate a minimal approach to analysing CP violation in this context using a Monte Carlo study of a combination of weak boson fusion Higgs and electroweak diboson production, which provide complementary information on the relevant EFT operators.
\end{abstract}


\author{Supratim~Das~Bakshi} \email{sdbakshi@iitk.ac.in}
\affiliation{Indian Institute of Technology Kanpur, Kalyanpur, Kanpur 208016, India\\[0.1cm]}
\author{Joydeep~Chakrabortty} \email{joydeep@iitk.ac.in}
\affiliation{Indian Institute of Technology Kanpur, Kalyanpur, Kanpur 208016, India\\[0.1cm]}
\author{Christoph~Englert} \email{christoph.englert@glasgow.ac.uk}
\affiliation{School of Physics \& Astronomy, University of Glasgow, Glasgow G12 8QQ, United Kingdom\\[0.1cm]}
\author{Michael~Spannowsky} \email{michael.spannowsky@durham.ac.uk}
\affiliation{Institute for Particle Physics Phenomenology, Department of Physics, Durham University, Durham DH1 3LE, United Kingdom\\[0.1cm]}
\author{Panagiotis~Stylianou}\email{p.stylianou.1@research.gla.ac.uk} 
\affiliation{School of Physics \& Astronomy, University of Glasgow, Glasgow G12 8QQ, United Kingdom\\[0.1cm]}

\preprint{IPPP/20/90}

\maketitle

\section{Introduction}
\label{sec:intro}
The lack of concrete evidence for new interactions beyond the Standard Model (BSM) 
has led to an increased consideration of effective field theory methods~\cite{Weinberg:1978kz} in high energy
collider physics. Besides their resurgence as largely model-independent interpretation
tools, EFT applications to collider searches and measurements have witnessed
rapid theoretical progress over the past years (see {\it e.g.}~Refs.~\cite{Brivio:2017vri,Dawson:2018dcd} for recent reviews). They 
increasingly become the {\emph{lingua franca}} adopted by the multi-purpose experiments
at the Large Hadron Collider (LHC) to report, and interpret results~(see {\it e.g.} the recent~\cite{Aad:2020sle}). While efforts are underway 
to extend SM effective field theory methods to even higher operator dimensions~\cite{Lehman:2015coa,Murphy:2020cly,Murphy:2020rsh,Li:2020gnx,Biekotter:2021int,Corbett:2021eux}, most analyses so far have focussed 
on deformations from dimension-six operator interactions~\cite{Burges:1983zg,Leung:1984ni,Buchmuller:1985jz,Hagiwara:1986vm,Grzadkowski:2010es} as leading terms in the effective field theory
expansion
\begin{equation}
\label{eq:eftd6}
{\cal{L}}_{\text{EFT}} = {\cal{L}}_{\text{SM}} + \sum_i {{\cal{C}}_i\over {\Lambda^2}} \, {\cal{O}}^{\text{d6}}_i\,.
\end{equation} 

Although the EFT methodology provides a theoretically consistent framework for the interpretation of particle physics measurements,
their generic approach to BSM searches typically, and unavoidably, leads to a significant shortfall in the new physics potential of
concrete LHC searches and measurements when many new interactions need to be considered. 
In the light of expected uncertainties and hadron collider energy coverage, the underlying EFT assumption of an absence of direct evidence of new propagating degrees of freedom often pushes constraints on the Wilson coefficients (WCs) ${{\cal{C}}_i/{\Lambda^2}}$ in Eq.~\eqref{eq:eftd6} to a regime where perturbative matching to concrete ultraviolet (UV) scenarios becomes challenged or impossible~(for a recent discussion see~\cite{Englert:2019rga}). This does not constitute a breakdown of EFT methods, but demonstrates a lack of a particular analysis' sensitivity to motivated UV completions of the SM, which is not visible from the constraint on the WCs themselves directly. 
While direct searches for motivated new degrees of freedom in concrete scenarios typically explore higher mass scales directly at hadron colliders compared to indirect deviations from expected SM correlations (see {\it e.g.}~\cite{Brown:2020uwk}), in parallel, concrete UV scenarios often also require a dedicated RGE-improved matching procedure see {\it e.g.}~Refs.~\cite{BuarqueFranzosi:2017jrj,Englert:2019rga}. More concretely: Specific UV scenarios impose hierarchies among WCs that are also reflected in their evolution over a broad range of momentum transfer scales that typically informs a LHC measurement result. While these questions arise beyond the mere application of EFT to collider data, they need to be taken into account when EFT methods are supposed to inform a more concrete UV picture, which are their {\emph{raison d`\^etre}}. 

Searches in the light of a large available EFT parameter space do always make assumptions, see e.g.~\cite{Aad:2020sle}. This raises the questions in how far these measurements can inform generic UV extensions, which is the purpose of this work. Indeed, we find that limiting the number of Wilson coefficients considered in asymmetry studies of CP-sensitive observables is theoretically well-motivated and justified. A general counter-argument that is raised for such investigations is that all information should be considered as part of a global statistical function such that matching computations are trivially includable. While this holds for naive extensions of the SM, this is less straightforward for loop-induced matching that contains logarithmic scale dependencies of physical processes~\cite{Englert:2019rga}. This would require experimentalists to track the full RGE flow effects as part of their investigation, which is currently not done and is unlikely to be included in the near future.

Starting from generic model considerations at the UV scale, we inform the EFT interpretation of TeV-scale measurements by landscaping WCs on the basis of anticipated UV scenarios' particle content and interactions. This enables us to reconcile and contextualise common assumptions with theoretically consistent UV completions, which is the aim of this work. We discuss the implications of generic UV scenarios for the particular subset of SMEFT operators that parametrise CP-violation in the gauge-Higgs sector. Building on previous work~\cite{DasBakshi:2020ejz}, we consider scalar and fermionic degrees of freedom to classify the operator patterns that arise from these theories in Sec.~\ref{sec:theory}, with a particular focus on the gauge representation of fermionic extensions. We gain evidence that generic analyses of CP violation by taking all {\emph{ad hoc}} EFT interactions into account on an equal footing (see, {\it e.g.}, the recent Refs.~\cite{Aad:2020mkp,Sirunyan:2020tqm,Sirunyan:2021zud,Aad:2020sle}) overestimate the expected phenomenological patterns of CP violation in the gauge-Higgs sector. The UV-EFT connection described in Sec.~\ref{sec:theory}, will therefore enable the LHC experimental collaborations to reduce their effective coupling parameter space in a theoretically well-motivated way, thus enhancing their analyses' sensitivities in ways that directly inform UV completions in a theoretically transparent fashion. In Sec.~\ref{sec:minfit}, we demonstrate such a procedure in a minimal fashion: The combination of weak boson fusion Higgs production, and mainly $W^\pm\gamma$ production combines complementary information when analysed with generically (C)P-sensitive observables (see~\cite{Plehn:2001nj,Hankele:2006ma,Klamke:2007cu,Campanario:2010mi,Brehmer:2017lrt,Bernlochner:2018opw,Englert:2019xhk,Cirigliano:2019vfc}). We conclude in Sec.~\ref{sec:conc}.

\section{Scenarios with low energy CP violation}
\label{sec:theory}
To start with, we consider that the SM particle content is extended by a heavy fermion $ \Psi $. The additional gauge-invariant renormalizable Lagrangian involving $ \Psi $ is written as
\begin{align}\label{generic-fermion}
		\mathcal{L}_{\Psi} &= \bar{\Psi}\left( i \slashed{D} - M_{\Psi}  \right) \Psi - \left\lbrace \bar{\psi}\, Y_{\psi}\, \Psi\,H + \text{h.c.} \right\rbrace ,
	\end{align}
where, the covariant derivative $ \slashed{D} $ is defined by the gauge quantum numbers of $ \Psi $ under the SM gauge symmetry $ SU(3)_C \otimes SU(2)_L\otimes U(1)_Y\, $.
The heavy fermion interacts with the SM fermions, and Higgs boson $ H $ through the Yukawa interactions, where $ \psi$ represents the light SM fermions, both quarks and leptons. Here, we are not displaying the flavour indices explicitly in Eq.~\eqref{generic-fermion}, as they are not relevant for our current discussion.  In principle, even if one invokes explicit CP-violation in these Yukawa couplings, these will contribute to an even number of $\gamma_{_5}$ in the spin traces leading to CP-conserving effective operators in the field strength $ (X^3) $ and the gauge-Higgs $ (\phi^2 X^2) $ classes.\footnote{These are dimension-six SMEFT Warsaw basis operators \cite{Grzadkowski:2010es}. The $ X^3 $ and $ \phi^2 X^2 $ refer to the triple-field-strength and the two-Higgs-two-field-strength Warsaw basis effective operator classes.}  In pursuit of CP-violating (CPV) effective operators, we require an odd number of $\gamma_{_5}$ matrices in the spin traces as such structures generate the fourth rank Levi-Civita tensor $ \epsilon_{\mu\nu\rho\sigma} $, {\it i.e.}, in turn generates the dual of the field strength tensors. In particular, this holds at loop-level~\cite{Jegerlehner:2000dz,Chanowitz:1979zu,Ahmed:2020kme,Heller:2020owb}. The single heavy fermion extension of the SM therefore fails to generate these CPV operators. Thus, in our next attempt, we add two heavy vector-like fermions to find out whether it is possible to generate the loop-induced CPV operators or not. 

\begin{figure}[!t]
	\includegraphics[height=10cm,width=15cm]{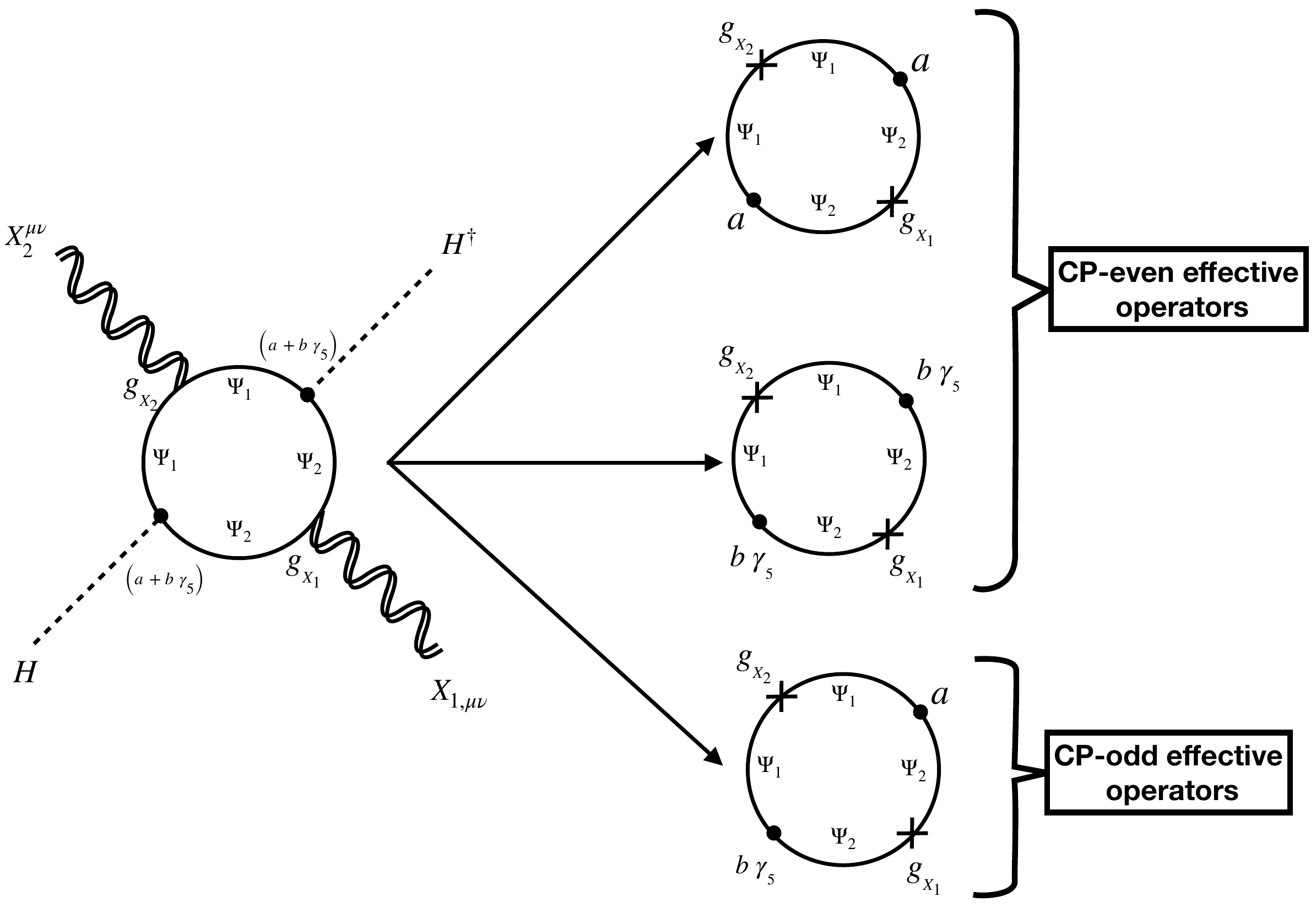}
	\caption{$X \, ( = B, W, G )\, $ represents the SM field strength tensors. For $ X_1=X_2  $, we get the SMEFT dimension-six operators $ Q_{HW}$, $ Q_{HB}$, $ Q_{HG}$, $ Q_{H\widetilde{W}}$, $ Q_{H \widetilde{B}} $ and $Q_{H\widetilde{G}}$, and for $ X_1 \neq X_2 $, we get the operators $ Q_{HWB} $ and $ Q_{H\widetilde{W}B} $. Depending on the number of $ \gamma_{_5} $s in the vertices, we categorise the diagram in the LHS into two cases: CP-even and CP-odd effective operator generating diagrams. The diagrams with an even number of $\gamma_{_5}$ vertices generate to CP-even operators and an odd number of $\gamma_{_5}$ vertices generate to CP-odd effective operators.}
	\label{fig:gauge-higgs}
\end{figure}
\begin{figure}[!t]
	\includegraphics[height=10cm,width=15cm]{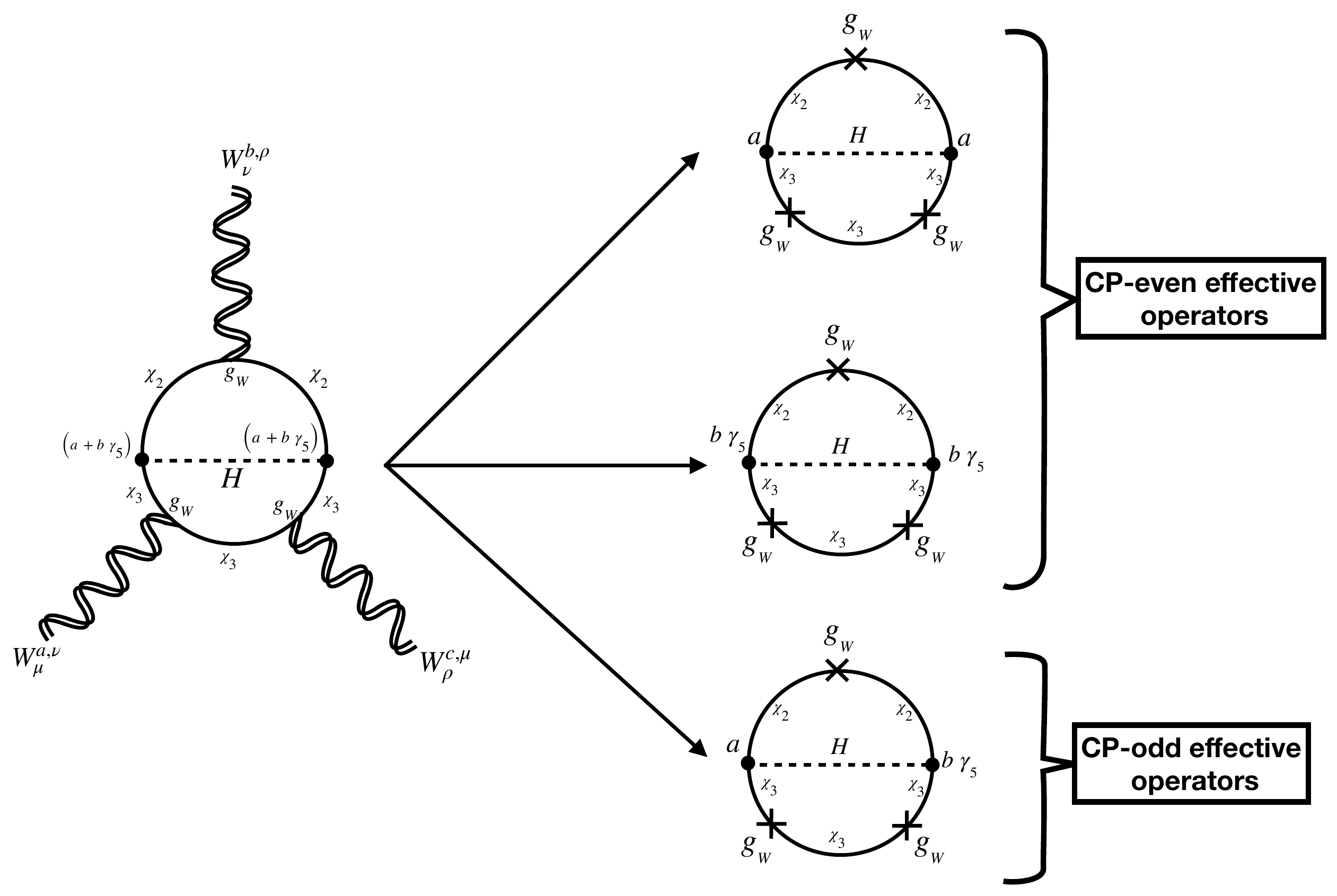}
	\caption{Two-loop diagram generating the SMEFT Warsaw basis $ Q_W $ and $ Q_{\widetilde{W}} $ operators by integrating out heavy VLL, see Eq.~\eqref{eq:DTVLL}. Unlike the $ Q_W $ operator which is generated at one-loop processes, the $Q_{\widetilde{W}} $ is generated at two-loop processes. The diagram in the bottom-right contributes to the WCs of the $ Q_{\widetilde{W}} $ operator. The CP-even and odd couplings are defined as $ a = \alpha_\chi$ and $ b=\beta_\chi $ in this model, see Eq.~\eqref{eq:defab}. We discuss more on the loop calculation and the coefficients of the $Q_{\widetilde{W}} $ operator in Appendix~\ref{sec:two-loop-calc}.}
	\label{fig:twoloopDT}
\end{figure}
The new  Lagrangian with two heavy vector-like leptons (VLL) $ \Psi_1 $ and $ \Psi_2 $ is written as
\begin{align}\label{eq:twoheavyfer}
	\mathcal{L}_{2\Psi} &= \bar{\Psi}_1\left( i \slashed{D} - M_{\Psi_1}  \right) \Psi_1 +\bar{\Psi}_2\left( i \slashed{D} - M_{\Psi_2}  \right) \Psi_2  - \left\lbrace  \bar{\Psi}_1 \, Y_\Psi\,  V  \, \Psi_2 + \text{h.c.} \right\rbrace,
\end{align}
where, $ V $ contains both CP-even and odd interactions, $ V = H \left(a \, \mathbb{1}_4 + b \, \gamma_{_5}\right) $ with `$ a $' and `$ b $' parametrising the CP-even and odd couplings, respectively. These couplings are complex in general, but we consider them as real in our discussion for simplicity without loss of any generality. The loop diagrams with an even number of $\gamma_{_5}$ vertices generate CP-even operators and an odd number of $\gamma_{_5}$ vertices generate CP-odd effective operators, see Fig.~\ref{fig:gauge-higgs}. The rank four Levi-Civita tensor $ \epsilon_{\mu\nu\rho\sigma} $ is induced using
\begin{align}
	2 i \sigma_{\rho\sigma} \gamma_{_5} &= \epsilon_{\mu\nu\rho\sigma} \sigma^{\mu\nu} ,
\end{align}
where, $ \sigma_{\mu\nu} = \frac{i}{2}\left[\gamma_\mu,\gamma_\nu\right] $ is  the Lorentz generator for the spin-$\frac{1}{2}$ fields. 

In this minimal scenario, the SM gauge quantum numbers of the heavy fermions are restricted. In general, the allowed gauge quantum numbers are: $ \Psi_1 \mapsto (\{\mathbb{1},R_C\},\{\mathbb{1},R_L\}+1,\{0,\mathcal{Y}\}+1/2)$ and $ \Psi_2 \mapsto (\{\mathbb{1},R_C\},\{\mathbb{1},R_L\},\{0,\mathcal{Y}\})$, where $R_C $, $ R_L $, and $ \mathcal{Y} $ are the quantum numbers under the SM gauge group $ SU(3)_C\otimes SU(2)_L\otimes U(1)_Y $ respectively. We discuss below one minimal example model where the SM is extended by an isospin-doublet and an isospin-triplet heavy VLL.

\begin{table*}[!htb]
	\caption{The one-loop generated $ X^3 $ and $ \phi^2 X^2 $ operators classes and their WCs after integrating out the heavy VLL in Eq.~\eqref{eq:DTVLL}. The  CPV  operators are displayed in first three rows. To compactify the result, we take the multiplicative factor $({16 \pi^2 m^2})^{-1}$ out of the WCs $ \left(\mathcal{C}_i\right) $, see Eq.~\eqref{eq:eft}.}
	\label{tab:warsaw}
	\vskip 0.2cm
	\centering
	\renewcommand{\arraystretch}{1.5}
	\begin{threeparttable}
		\begin{tabular}{|c|c|l|}
			\hline\hline 
			$\text{Operators}$&$\text{Operator Structures}$&$\text{Wilson coefficients }(\mathcal{C}_i)$\\
			\hline\hline
			$Q_{H\widetilde{B}}$  &  $\left(H^{\dagger }H^{  }\right)\widetilde{B}_{\mu \nu }B^{\mu \nu }$  &   $-4g_Y^2 \text{Im}\left[y_{\chi_{_L}} y^\ast_{\chi_{_R}} \right]$  \\
			\hline 
			$Q_{H\widetilde{W}}$  &  $\left(H^{\dagger }H^{  }\right)\widetilde{W}_{\mu \nu }{}^IW^{I,\mu \nu }$  &  $- g_{_W}^2 \text{Im}\left[y_{\chi_{_L}} y^\ast_{\chi_{_R}} \right]$  \\
			\hline
			$Q_{H\widetilde{W}B}$  &  $\frac{1}{2}\left(H^{\dagger }\sigma ^I H^{  }\right)\widetilde{W}_{\mu \nu }{}^I B^{\mu \nu }$  & $-\frac{10}{3} g_{_W} g_{_Y}  \text{Im}\left[y_{\chi_{_L}} y^\ast_{\chi_{_R}} \right] $ \\
			\hline
			\hline
			$Q_{{HB}}$&  $\left(H^{\dagger }H^{  }\right)B_{\mu \nu }B^{\mu \nu }$  &
			$\frac{1}{60}g_Y^2\left[19|\alpha_\chi|^2 +15 |\beta_\chi|^2 \right]$  \\
			\hline
			$Q_{{HW}}$  &  $\left(H^{\dagger }H^{  }\right)W_{\mu \nu }{}^I W^{I,\mu \nu }$  &  $\frac{1}{6}g_{_W}^2\left[|\alpha_\chi|^2+|\beta_\chi|^2\right]$  \\
			\hline
			$Q_{{HWB}} $  &  $\frac{1}{2}\left(H^{\dagger }\sigma ^I H^{  }\right)W_{\mu \nu }{}^I B^{\mu \nu }$  &  $\frac{2}{15} g_{_W} g_{_Y}\left[6|\alpha_\chi|^2 + 5|\beta_\chi|^2\right]$  \\
			\hline
			$ Q_{W} $ & $ \epsilon^{IJK} W^{I\nu}_{\mu} W^{J \rho}_{\nu} W^{K\mu}_{\rho} $ & $\frac{7}{180}g_{_W}^3$ \\
			\hline \hline
		\end{tabular}
	\end{threeparttable}
\end{table*}

\begin{table*}[!htb]
	\caption{The one-loop generated dimension-eight CPV $ \phi^4 X^2 $ operators and their WCs after integrating out the heavy VLL in Eq.~\eqref{eq:DTVLL}. These operators modify the contributions of the dimension-six CPV operators shown in Tab.~\ref{tab:warsaw}.  These SMEFT dimension-eight operator structures are taken from Ref.~\cite{Murphy:2020rsh}.}
	\label{tab:dim8cpodd}
	\vskip 0.2cm
	\centering
	\renewcommand{\arraystretch}{1.5}
	\begin{threeparttable}
		\begin{tabular}{|c|c|c|}
			\hline\hline 
			$\text{Operators}$&$\text{Operator Structures}$&$\text{Wilson coefficients }(\mathcal{C}_i) \times ({16 \pi^2 m^4})^{-1}$\\
			\hline\hline
			$Q^{(2)}_{B^2H^4}$  &  $\left(H^{\dagger }H^{  }\right)^2\widetilde{B}_{\mu \nu } B^{\mu \nu }$  &   $- \frac{5}{3} g_{_Y}^2  \left[|\alpha_\chi|^2 -2 |\beta_\chi|^2\right]  \text{Im}\left[y_{\chi_{_L}} y^\ast_{\chi_{_R}} \right]$  \\
			\hline
			$Q^{(2)}_{W^2H^4}$  &  $\left(H^{\dagger }H^{  }\right)^2\widetilde{W}_{\mu \nu }{}^IW^{I,\mu \nu }$  &   $ -\frac{1}{10}g_{_W}^2  \left[2|\alpha_\chi|^2 -5 |\beta_\chi|^2\right]  \text{Im}\left[y_{\chi_{_L}} y^\ast_{\chi_{_R}} \right]$  \\
			\hline
			$Q^{(2)}_{WBH^4}$  &  $\frac{1}{2}\left(H^{\dagger }H^{  }\right) \left(H^{\dagger } \sigma^I H^{  }\right) \widetilde{W}^{I,\mu \nu } B_{\mu \nu }$   &   $- \frac{16}{15}   g_{_W} g_{_Y}  \left[|\alpha_\chi|^2 -5 |\beta_\chi|^2\right] \text{Im}\left[y_{\chi_{_L}} y^\ast_{\chi_{_R}} \right]$  \\
			\hline\hline
		\end{tabular}
	\end{threeparttable}
\end{table*}

\begin{table*}[!htb]
	\caption{Ratio of operators and Wilson coefficients at dimension-eight and -six.}
	\label{tab:dim8-6ratio}
	\vskip 0.2cm
	\centering
	\renewcommand{\arraystretch}{2.8}
		\begin{tabular}{|c|}
			\hline\hline 
			$\frac{\mathcal{C}^{(2)}_{B^2H^4}}
			{\mathcal{C}_{H\widetilde{B}}
			} = \left.\frac{(H^\dagger H)}{m^2} \times\frac{ (- \frac{5}{3} \left[|\alpha_\chi|^2 -2 |\beta_\chi|^2\right] )}{-4}\right\rvert_{\langle H^\dagger H \rangle = \frac{v^2}{2}} = \frac{v^2}{2m^2} \times\frac{5 \left[|\alpha_\chi|^2 -2 |\beta_\chi|^2\right]}{12} $\\
			$\frac{\mathcal{C}^{(2)}_{W^2H^4}}
			{\mathcal{C}_{H\widetilde{W}}}
			= \left.\frac{(H^\dagger H)}{m^2} \times\frac{(-\frac{1}{10}  \left[2|\alpha_\chi|^2 -5 |\beta_\chi|^2\right] )}{-1}\right\rvert_{\langle H^\dagger H \rangle = \frac{v^2}{2}} = \frac{v^2}{2m^2} \times\frac{\left[2|\alpha_\chi|^2 -5 |\beta_\chi|^2\right]}{10}  $\\
			$\frac{\mathcal{C}^{(2)}_{WBH^4}}
			{\mathcal{C}_{H\widetilde{W}B}}
			= \left.\frac{(H^\dagger H)}{m^2} \times\frac{{(- \frac{16}{15}     \left[|\alpha_\chi|^2 -5 |\beta_\chi|^2\right])}}{-\frac{10}{3}}\right\rvert_{\langle H^\dagger H \rangle = \frac{v^2}{2}} = \frac{v^2}{2m^2} \times\frac{{8\left[|\alpha_\chi|^2 -5 |\beta_\chi|^2\right]}}{25}  $\\
			\hline\hline
		\end{tabular}
\end{table*}


\begin{figure}[]
\centering
\parbox{0.5\textwidth}{{\includegraphics[width=0.48\textwidth]{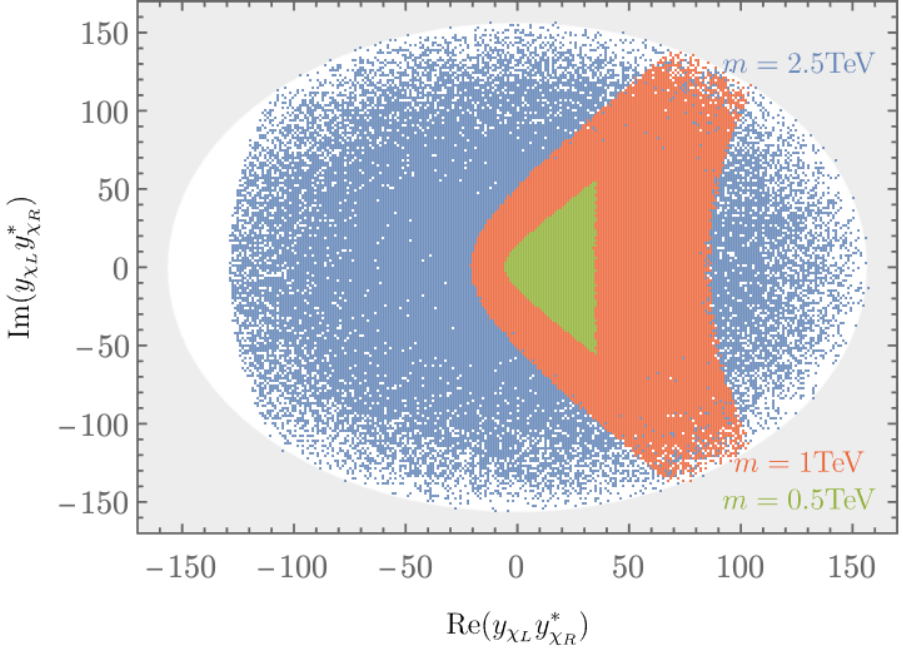}}}
\hspace{1cm}\parbox{0.38\textwidth}{
	\caption{Points in the $\Reylyr$-$\Imylyr$ plane were the leading contribution arises from dimension-six operators. The scan was performed for three different VLL masses by sampling $y_{\chi_L}$ and $y_{\chi_R}$ such that the absolute values of the ratios shown in Tab.~\ref{tab:dim8-6ratio} are all less than unity. Shaded with gray is the non-perturbative region where $\lvert y_{\chi_{L,R}} \rvert > 4 \pi$. \label{fig:dim8scan}}}
\end{figure}

\subsection{Doublet-Triplet VLL}
We consider two heavy VLL $ \chi_{_2} $ and $ \chi_{_3} $ with masses $ m_{\chi_{_2}} $ and $ m_{\chi_{_3}} $ respectively where their SM gauge quantum numbers are
\begin{equation}\label{eq:heavyfield-qnos}
\left(\chi_{_2}\right)_{L,R} : (1,2,1/2),\;\;
\left(\chi_{_3}\right)_{L,R} : (1,3,1).
\end{equation}
As explained above, these heavy fields interact with the SM Higgs doublet through the Yukawa interactions that accommodate the explicit CP violations leading to the generation of CPV operators. The relevant part of the BSM Lagrangian, containing these heavy leptons, is written as
\begin{multline}\label{eq:DTVLL}
\mathcal{L}_{\text{DT}}  \supset  \bar{\chi}_{_2} ( i \D_{\chi_{_2}} - m_{\chi_{_2}}) \chi_{_2} + \bar{\chi}_{_3} ( i \D_{\chi_{_3}} - m_{\chi_{_3}}) \chi_{_3} 
- \left\lbrace H^T \, \bar{\chi}_{_3}^I\, \sigma^I  ( y_{\chi_{_L}} \mathbb{P}_L + y_{\chi_{_R}} \mathbb{P}_{R}) i \sigma_2\, \chi_{_2}+ \text{h.c.} \right\rbrace.
\end{multline}
Here, $ I\, (= 1,2,3) $ is the $ SU(2)_L $ index of the isospin-triplet lepton $ \chi_{_3} $, the $ \sigma^I $ represent the Pauli matrices, and $ \mathbb{P}_L (\mathbb{P}_R)$ is the left(right) chiral projection operator. Integrating out these heavy VLL and matching this BSM to the SMEFT, we determine the WCs of the $ X^3 $ and $ \phi^2 X^2 $ operator classes. The effective Lagrangian is defined as 
\begin{align}\label{eq:eft}
\mathcal{L}_{\text{EFT}}= \mathcal{L}_{\text{SM}} + \frac{1}{16 \pi^2 m^2} \sum\limits_i  {\mathcal{C}}_i Q_i \,,
\end{align}
where $ m $ is the cut-off scale of the effective theory. The masses of the VLL are taken as degenerate ($ m = m_{\chi_{_2}}=m_{\chi_{_3}} $). 
 We note that the heavy fermions are $ SU(3)_C $ singlet fields in this example model. Therefore, the following effective operators belonging to $X^3$ and $\phi^2\,X^2$ classes are generated after matching:
\begin{equation}\label{eq:opsgenDT}
\left\lbrace 	Q_{W},  Q_{HW}, Q_{HB}, Q_{HWB} 	\right\rbrace \cup 
\left\lbrace  Q_{\widetilde{W}},Q_{H\widetilde{W}},Q_{H\widetilde{B}},Q_{H\widetilde{W}B} \right\rbrace
\, .
\end{equation}
 We tabulate the matched WCs associated with these operators in Tab.~\ref{tab:warsaw}, where we have defined
 	\begin{align}\label{eq:defab}
 		\alpha_\chi = \frac{y_{\chi_{_R}} + y_{\chi_{_L}}}{2} \, , \ \ \text{and} \ \  \beta_\chi= \frac{y_{\chi_{_R}} - y_{\chi_{_L}}}{2},
 	\end{align}
	 to obtain compact expressions of the WCs. Here, we assume that the left $ (y_{\chi_{_L}}) $ and the right $ (y_{\chi_{_R}}) $ chiral couplings are different. This is necessary to generate the CPV operators, because the WCs of these operators are proportional to their difference, as shown in Tab.~\ref{tab:warsaw}.  
It is worthwhile to mention that if we consider coloured heavy fermions, the additional operators
\begin{equation}\label{eq:opsgencolorBSM}
\left\lbrace 	Q_{G}, Q_{HG} \right\rbrace \cup \left\lbrace Q_{\widetilde{G}},  Q_{H\widetilde{G}} \right\rbrace 
\end{equation}
are also generated. 

We integrate out the heavy fermions using a framework work based on the functional methods developed in Ref.~\cite{Angelescu:2020yzf} to calculate these WCs at one-loop. As this will be important in the following, we emphasise that the $ Q_{\widetilde{W}} $ operator does not arise at one-loop-level, and we have to look beyond one-loop matching to generate this operator. In Fig.~\ref{fig:twoloopDT}, we show a schematic diagram depicting the origin of this particular CPV operator in this model at two-loop, which is also the leading order contribution at dimension-six. We note that these processes get contributions from other higher dimension operators generated at 1-loop, for example, the $ \phi^2 X^3 $-class of dimension-eight operators, after $\phi$ acquires $vev$. The three relevant CPV dimension-eight operators are shown in Tab.~\ref{tab:dim8cpodd} along with the matching of their respective WCs. Tab.~\ref{tab:dim8-6ratio} shows the ratio of their contributions with respect to dimension-six when the Higgs doublet acquires the $vev$ and in order to identify the regions where higher orders are suppressed we perform the scan over $y_{\chi_L}$ and $y_{\chi_R}$ of Fig.~\ref{fig:dim8scan}. For perturbative couplings of a heavy VLL with mass $m \gtrsim 2.7$~TeV, dimension-six is dominant, while for smaller masses the same applies for certain regions depending on the value of $\Reylyr$. It is worth highlighting that when EFT is a valid approach for comparably large VLL masses, the validity of the dimension-six approach is set by the perturbativity of the couplings alone. This is due to the fact that the dimension-eight operators of Tab.~\ref{tab:dim8cpodd} are sensitive to the same coupling combination as the dimension-six matching. In a scenario, where one considers contributions from dimension-eight or higher dimensional operators, the dominant contributions to the WCs of these processes will be determined by the interplay of the loop suppressions  $ versus $ the cut-off suppression of the dimension-eight or more. We elaborate on the contributions of these diagrams to the WCs of $ Q_{\widetilde{W}} $ operator in Appendix~\ref{sec:two-loop-calc}, where we also comment on the non-appearance of $Q_{\widetilde{W}} $ in scalar extensions of the SM. 

\section{A minimal fit of CP violation}
\label{sec:minfit}
In the light of the previous section, we can now outline the approach to capture a complete picture of CP violation at dimension-six level. With $ Q_{\widetilde{W}} $ absent at one-loop matching order in realistic and perturbative extensions of the SM, CP violation related to electroweak interactions can be dissected through a combination of diboson and electroweak Higgs measurements, and in particular observed asymmetries of tailored CP-sensitive distributions in these channels~\cite{Miller:2001bi,Choi:2002jk,Accomando:2006ga,Bernlochner:2018opw,Englert:2019xhk}. These observables also have the benefit of being insensitive to CP-even deformations of the SM interactions, in particular to ``squared'' dimension-six CP-odd contributions. This makes asymmetries particularly well-suited to pinpoint CP violation, while the CP-even cross section and rate information enters the limit setting through statistical uncertainties.

We will outline that possible CP violation introduced via the $Q_{H\widetilde{W}}$, and $Q_{H\widetilde{B}}$ operators can be efficiently captured by the differential distribution of sensitive observables in Higgs production through Weak Boson Fusion (WBF) channels~\cite{Barger:1999tn,Plehn:2001nj,Hankele:2006ma,Englert:2012xt,Englert:2012ct, Dolan:2014upa, Araz:2020zyh,Ethier:2021ydt}, as well as diboson production~\cite{Franceschini:2017xkh,Banerjee:2019twi,Banerjee:2020vtm}, predominantly $pp\to W\gamma$~\cite{DasBakshi:2020ejz,Biekotter:2021int}. 

\subsection{WBF Analysis}
We study the effects of the additional operators in WBF production $p p \to H j_1 j_2$, concentrating on Higgs decays to $\tau$ leptons only \cite{Dolan:2014upa, Araz:2020zyh}, due to the challenging nature of $H \to b \bar{b}$~\cite{Khachatryan:2015bnx,Englert:2015dlp,Ngairangbam:2020ksz}, although additional channels should significantly enhance sensitivity \cite{Andersen:2012kn}. Our toolchain consists of {\sc{FeynRules}}~\cite{Christensen:2008py,Alloul:2013bka} for modelling the BSM operators which produces a {\sc{Ufo}}~\cite{Degrande:2011ua} model file. Through the {\sc{MadGraph}}~\cite{Alwall:2014hca} framework, the model is imported for event generation using {\sc{MadEvent}}~\cite{Alwall:2011uj,deAquino:2011ub,Alwall:2014hca}. Events are generated with a Higgs mass of $125$~GeV and total decay width $\Gamma_H = 4.2$~MeV, and we perform our analysis by imposing selection criteria at parton-level.

Large rapidity separation $y$ and back-to-back jets with large invariant-mass are characteristics of the WBF signal topology~\cite{Dokshitzer:1987nc,Dokshitzer:1991he,Barger:1991ar,Bjorken:1992er}, commonly used for discrimination from contributing backgrounds. We thus require at least two jets with $p_T > 25$~GeV and invariant mass of the two leading jets of $m_{jj} > 800$~GeV. Additionally, the search region is constrained imposing a cut on the absolute rapidity difference $\abs{\Delta y_{jj}} > 2.8$. 

We define the efficiency of identifying the Higgs as the ratio of the SM cross section rates of $p p \to (H \to \tau \bar{\tau}) j_1 j_2$ with tagged taus divided by the rate of $p p \to H j_1 j_2$ with identical jet cuts. Higgs candidates are identified by requiring two $\tau$-tagged leptons with $p_T(\tau) > 40$~GeV and isolated from each other\footnote{We define isolation of the two taus by requiring $\Delta R = \sqrt{(\Delta \eta)^2 + (\Delta \phi)^2} > 0.4$, where $\Delta \eta$ and $\Delta \phi$ are the pseudorapidity and azimuthal separations between them.}, such that their reconstructed invariant mass is within $10$~GeV of the Higgs. Taus are identified with an efficiency for purely-hadronic decay of 85\% and a branching ratio $BR(\tau \to \text{hadrons}) = 0.65$. Our final Higgs identification efficiency was subsequently calculated as 1.03\% and is used to rescale the event samples of $p p \to H j_1 j_2$ generated with and without the introduction of the additional operators. Finally, the parity-sensitive signed azimuthal angle $\Delta \phi_{j_1 j_2}$ is calculated for the events satisfying our search requirements.

\subsection{Diboson Analysis}
The analysis of diboson final states of Refs.~\cite{DasBakshi:2020ejz,Biekotter:2021int} complements the WBF search detailed above. In particular, parity-sensitive signed azimuthal angles are constructed from final states of $W\gamma \to \ell \nu \gamma$, $W^+ W^- \to \ell^+ \nu_\ell \ell^- \bar{\nu}_\ell$ and $W Z \to \ell \nu \ell^+ \ell^-$ processes, modified by the introduction of the three CP-violating operators.\footnote{We use the same toolchain as the WBF analysis, unless stated otherwise.}

\subsubsection*{$W\gamma \to \ell \nu \gamma$ process}
The decays of $W$ in the $W\gamma$ production to electrons and muons are considered separately with different cuts applied following Ref.~\cite{CMS:2020olm}. Events are generated with zero and one jet, and are merged using the MLM scheme through {\sc{Pythia8}}~\cite{Sjostrand:2014zea}.\footnote{Including additional jet emission in diboson production is essential to reflect sensitivity-limiting hard jet emission, which impacts, {\it e.g.}, the radiation zero, see below. Additional jet radiation is a not as relevant for WBF type selections where jet emission follows a bremstrahlung paradigm, with little impact on the leading jet correlations themselves~\cite{Figy:2003nv}.}

 We use the reconstruction mode of {\sc{MadAnalysis}}~\cite{Conte:2012fm,Conte:2014zja,Dumont:2014tja,Conte:2018vmg} to interface {\sc{FastJet}}~\cite{Cacciari:2011ma,Cacciari:2005hq} and cluster jets using the anti-kT algorithm~\cite{Cacciari:2008gp} of radius $0.4$. The search region is constraint to events that have a total missing energy $\slashed{E}_T > 40$, arising from the neutrino of the $W$ boson decay. Exactly one isolated electron (muon)\footnote{Electrons and muons are considered isolated if the sum of transverse momenta of all jets within $\Delta R < 0.4$ is less than 50\% of the lepton's $p_T$.} is required with absolute pseudorapidity $\abs{\eta_e} < 2.5$ ($\abs{\eta_\mu} < 2.4$) and transverse momentum $p_T(e) > 30$~GeV ($p_T(\mu) > 26$~GeV) and no other lepton with $p_T > 20$~GeV. Additionally, at least one isolated photon\footnote{Isolated photons satisfy the same criteria as isolated electrons and muons.} separated from the lepton by $\Delta R (\gamma, \ell) > 0.5$ and satisfying $\abs{\eta} < 2.5$,  $p_T(\gamma) > 25$~GeV must be identified. We define a rate normalisation factor for electrons and muons as $R^{e,\mu} = N_{\text{CMS}}^{e,\mu} / N_{\text{MC}}^{e,\mu}$, where $N_{\text{CMS}}^{e} = 96000$ and $N_{\text{CMS}}^\mu = 164400$ are the number of events observed in the signal region by CMS~\cite{CMS:2020olm} based on an integrated luminosity of $137.1$/fb, and $N_{\text{MC}}^{e,\mu}$ is the number of events we obtain after the aforementioned cuts without the effects of the BSM operators. The signed azimuthal angle in this particular process is then defined as $\Delta \phi_{\gamma \ell} = \phi_{\gamma} - \phi_{\ell}$ ($\Delta \phi_{\gamma \ell} = \phi_{\ell} - \phi_{\gamma}$) if the photon's rapidity is greater (less) than the lepton's. It should be noted that although at LO the $\Delta \phi_{\gamma \ell}$ distribution has two distinct and well-separated peaks; additional jet activity significantly distorts this behaviour leading to a considerable reduction of sensitivity. This is in line with the absence of the so-called radiation zero~\cite{Samuel:1983eg,Samuel:1984ru,Brodsky:1982sh,Brown:1982xx} of $W\gamma$ production when additional jet emission is considered~\cite{Diakonos:1992qc,Baur:1993ir,Campanario:2010hv}. It is therefore beneficial in this particular search to impose an additional veto on jets with $p_T(j) > 30$~GeV to enhance the sensitivity to the BSM couplings~\cite{Baur:1993ir}.

\subsubsection*{$W^+ W^- \to \ell^+ \nu_\ell \ell^- \bar{\nu}_\ell$ process}
We include the $WW$ channel in the diboson analysis by considering the $e \nu_e \mu \nu_\mu$ final state as in Ref.~\cite{Aaboud:2019nkz} and subsequently rescaling to include final states for all the light lepton modes of the channel. A requirement of $E_T > 20$~GeV is imposed on the missing energy and exactly one electron and one muon must be identified with $p_T > 10$~GeV with no jet activity in the cone $\Delta R < 0.4$. Additional cuts $\abs{\eta(\ell)} < 2.5$ and $p_T(\ell) > 27$~GeV are then imposed on the two leptons. The reconstructed invariant mass and transverse momentum of the dilepton constrain the search region via enforcing $m_{e\mu} > 55$~GeV and $p_T(e\mu) > 30$~GeV, respectively, which suppress the Drell-Yan and $H\to WW$ background contamination. The rate of the accepted events is normalised to the fiducial cross section of the channel $\sigma_\text{fid} = 379.1$~fb, calculated by ATLAS~\cite{Aaboud:2019nkz}. The observable studied in this case is $\Delta \phi_{\ell\ell}$, where as in the other channels the order that the azimuthal angles are subtracted depends on the rapidity ordering of the leptons.

\subsubsection*{$W Z \to \ell \nu \ell^+ \ell^-$ process}
The $WZ$ channel is characterised by one same-flavor opposite-charge lepton pair originating from the $Z$ boson and an additional lepton. Thus, exactly three leptons should appear in the region $\abs{\eta(\ell)} < 2.5$ and $p_T(\ell) > 5$~GeV without jets within small separation $\Delta R < 0.4$ for an event to be considered. Furthermore, we reconstruct the invariant mass of the candidate pair with opposite charges satisfying $61 \leq m_{\ell\ell} \leq 121$ (in GeV) and consider the remaining lepton as the one originating from the $W$ boson. The search region is constrained by imposing a cut on the $W$-induced lepton $p_T(\ell) > 20$~GeV. We define the signed azimuthal angle $\Delta \phi_{\ell Z}$ by reconstructing the $Z$ boson's rapidity and azimuthal angle (based on the four-momentum calculated as the sum of the dilepton pair 4-momenta). The rate of the process is then normalised to the fiducial cross section of this specific search region obtained by CMS, $\sigma_{\text{fid}} = 258$~fb~\cite{Khachatryan:2016tgp}.

\subsection{Statistical analysis of $\Delta \phi$ distributions}
\begin{figure}[!t]
\centering
\parbox{0.5\textwidth}{{\includegraphics[width=0.48\textwidth]{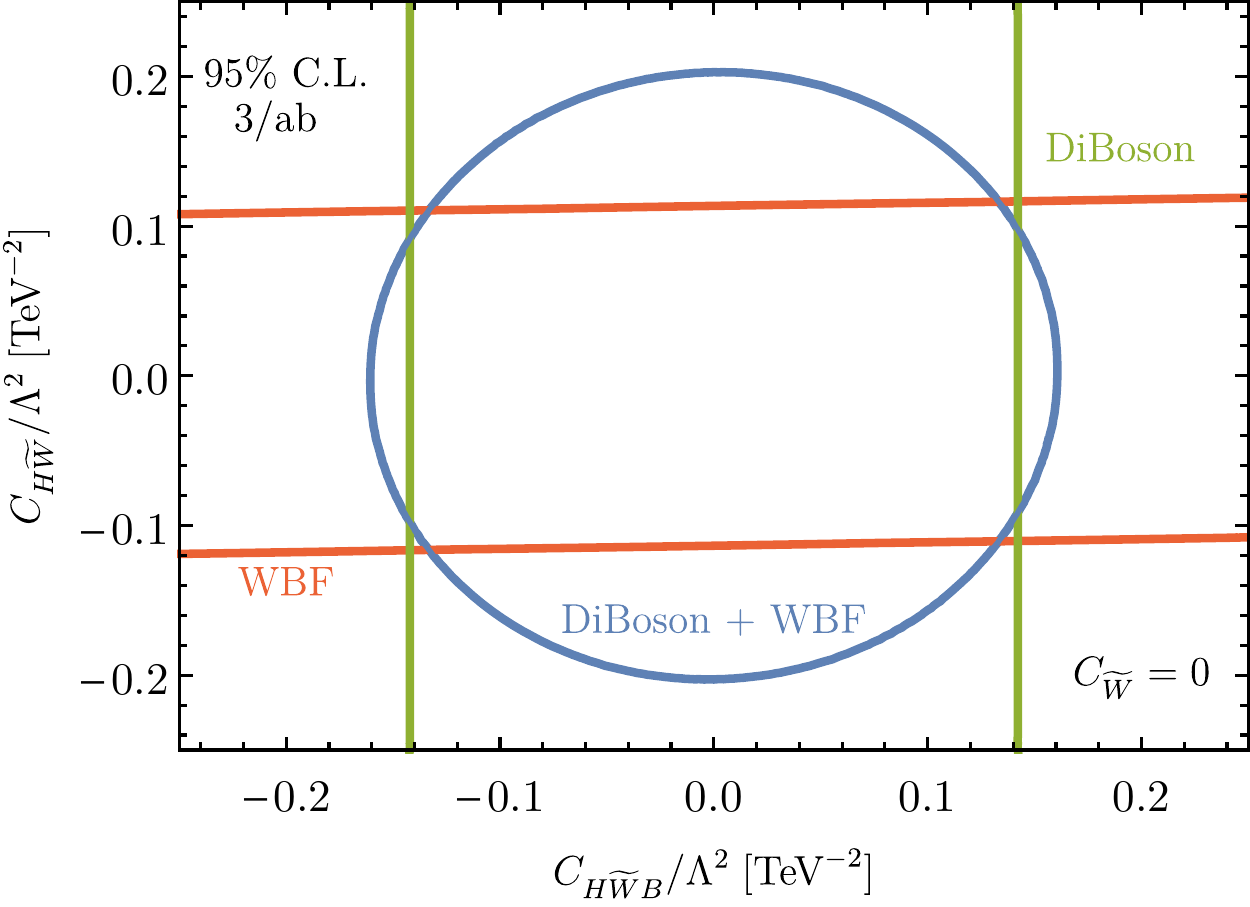}}}
\hspace{1cm}\parbox{0.37\textwidth}{
	\caption{Contours at 95\% C.L. for WBF analysis with the Higgs detected through the $h \to \tau^- \tau^+$ channel combined with di-boson analyses. Bounds are shown with $C_{\widetilde{W}}$ fixed at zero and we profiled over $C_{H\widetilde{B}}$.\label{fig:contours}}}
\end{figure}

For each of the studied processes, the differential distribution for $\Delta \phi_{ij}$ can be obtained, which includes linear contributions from the additional operators
\begin{equation}
	\label{eq:diffdist}
	\frac{\tx{d}\sigma(\chwb/\Lambda^2,\chw/\Lambda^2,\chb/\Lambda^2)}{\tx{d}\Delta \phi_{ij}} = \frac{\tx{d}\sigma_{\tx{SM}}}{\tx{d}\Delta \phi_{ij}} + \chwblam \frac{\tx{d}\sigma_{H\widetilde{W}B}}{\tx{d}\Delta \phi_{ij}} + \chwlam \frac{\tx{d}\sigma_{H\widetilde{W}}}{\tx{d}\Delta \phi_{ij}} + \chblam \frac{\tx{d}\sigma_{H\widetilde{B}}}{\tx{d}\Delta \phi_{ij}} \,.
\end{equation}
Cross sections $\sigma_{H\widetilde{W}B}$, $\sigma_{H\widetilde{W}}$ and $\sigma_{H\widetilde{B}}$ capture the interference effects introduced by the operators $Q_{H\widetilde{W}B}$, $Q_{H\widetilde{W}}$ and $Q_{H\widetilde{B}}$, respectively, but are independent by construction to their respective WCs. This allows us to scan various values of the three coefficients and by rescaling linearly we obtain different differential distributions of \eqref{eq:diffdist}. Subsequently, a $\chi^2$-statistic is calculated as 
\begin{equation}
	\label{eq:chi2}
	\chi^2\left(\chwblam,\chwlam,\chblam\right) = \sum_{i\in \text{bins}}\frac{(b_\tx{SM+d6}^i(\chwb/\Lambda^2,\chw/\Lambda^2,\chb/\Lambda^2) - b_\tx{SM}^i)^2}{\sigma_i^2}\,,
\end{equation}
where $b_\tx{SM+d6}^i(\chwb/\Lambda^2,\chw/\Lambda^2,\chb/\Lambda^2)$ denotes the number of events in the $i$-th bin, as obtained by multiplying \eqref{eq:diffdist} with an integrated luminosity, and $b^i_{\tx{SM}} = b_\tx{SM+d6}^i(0,0,0)$. The uncertainties $\sigma_i = \sqrt{b^i_{\tx{SM}}}$ are statistical fluctuations. Systematic errors could in principle be introduced via a covariance matrix, however in this particular setting, cancellations between the symmetric SM and antisymmetric BSM contributions in the $\chi^2$ sum would render them irrelevant. Confidence level intervals can be then calculated using
\begin{equation}
	\label{eq:cl}
	1 - CL \geq \int_{\chi^2}^{\infty}  {\text{d}}x \,p_k(x)\,,~\quad \chi^2=\chi^2\left(\chwblam,\chwlam,\chblam\right)\,,
\end{equation}
where $p_k(x)$ is the $\chi^2$ distribution of $k$ degrees of freedom evaluated by subtracting the number of fitted WCs from the number of total bins. A scan is performed with the $\chi^2$-statistic calculated at an integrated luminosity of $3$/ab, and contours evaluated in the $\chwb$-$\chw$ plane at 95\% confidence level for the WBF process and the combined diboson processes are shown in Fig.~\ref{fig:contours}. Contours are calculated by both fixing $\chb = 0$ and by profiling, where the $\chb$ value is determined such that it minimises the $\chi^2$ function.

\subsection{Discussion}
The result of the diboson-Higgs combination is  shown in Fig.~\ref{fig:contours}, for a LHC luminosity of 3/ab for different WC choices. As can be seen, these channels serve to constrain the dominant operator directions $\sim C_{\widetilde{W}},C_{H\widetilde{W}B}$ when $C_{\widetilde{W}}=0$ is injected into the fit.\footnote{This situation is somewhat similar to imposing $U=0$ in electroweak oblique corrections constraints~\cite{Peskin:1991sw,Peskin:1990zt} (see {\it e.g.} the fit reported in Ref.~\cite{Baak:2014ora}). While $U$ is not related to a dimension-six operator, $C_{\widetilde{W}}$ is two-loop suppressed for the scenarios considered in this work. When a perturbative matching is possible, a hierarchy $0\simeq C_{\widetilde{W}}$ as part of the set of Eq.~\eqref{eq:opsgenDT} is expected.} It is worth noting that the $C_{H\widetilde{B}}$ direction is difficult to constrain in the production and decay modes that we consider, predominantly because $C_{H\widetilde{B}}$ reflects weaker $U(1)_Y$ dynamics. In concrete scenarios, these are typically  suppressed compared to the $C_{H\widetilde{W}}$ direction. Our analysis is not sensitive to the $C_{H\widetilde{B}}$ direction as can be seen from the profiled contours of Fig.~\ref{fig:contours}, which shows that constraints on the expected dominant deviations from the SM can be obtained. Subsidiary measurements that specifically target this operator are given by $H\to \gamma\gamma$ and decay plane investigations of $H\to ZZ \to 4 \ell$ which then enhance the sensitivity to a level that constraints can be formulated on all contributing operators as demonstrated in Ref.~\cite{Bernlochner:2018opw}. In this sense, the inclusion of more Higgs data will further enhance the sensitivity, however, an inclusive picture of the CP-violation can be obtained from the combination of Higgs and diboson data. When the results of Fig.~\ref{fig:contours} are used to inform a concrete UV scenario using a matching calculation, the $C_{H\widetilde{W}}$, $C_{H\widetilde{W}B}$ constraints are sufficient to constrain the parameter space as $C_{H\widetilde{B}}$ is related through the gauge representation. Additional sensitivity to $C_{H\widetilde{B}}$ would then reflect the hypercharge assignments which can also be obtained from CP-even operator measurements~\cite{DasBakshi:2020ejz}.
Since our analysis does not include contributions from dimension-eight, its validity applies only when the higher order deformations are suppressed compared to the included dimension-six. Concerns of the sensitivity of concrete analyses when phrased as EFT constraints have raised questions concerning the self-consistency of EFT approaches, see, e.g., Refs.~\cite{Kalinowski:2018oxd,Kozow:2019txg,Lang:2021hnd} and \cite{ATLAS:2016snd,CMS:2020gfh}. Coming back to the discussion in Sec.~\ref{sec:theory} (see also Tab.~\ref{tab:dim8-6ratio} and Fig.~\ref{fig:dim8scan}) we can address this potential issue through our dimension-six and -eight matching calculation. When interpreted in terms of a heavy VLL, our constraints lie in the range of values for the WCs where contributions from dimension-six are indeed larger. The points of Fig.~\ref{fig:dim8scan} for $m = 1$~TeV (such mass scales are outside the LHC coverage, see e.g.~\cite{Buchmueller:2013dya}) result in WCs in the regions $\lvert C_{H\widetilde{B}}\rvert \lesssim 0.4$, $\lvert C_{H\widetilde{W}}\rvert \lesssim 0.4$ and $\lvert C_{H\widetilde{W}B}\rvert \lesssim 0.6$.

\section{Summary and Conclusions}
\label{sec:conc}
Effective field theory is rapidly (and deservedly) becoming the lingua franca of disseminating hadron collider measurements and commenting on these measurements' agreement or tension with the SM expectation. Ultimately, EFT methods should inform concrete scenarios that extend the SM towards higher energies. The agnostic approach of the generic dimension-six extension of the SM~\cite{Grzadkowski:2010es} can be misleading in this regard. Concrete scenarios will exhibit hierarchical WC structures due to their particle and symmetry content, which need to be included to enable a translation of EFT constraints into relevant model parameter regions. This selects regions between the fully marginalised constraints and by-hand zero WC choices, which are the two customary avenues to report results. 

In this work, we approach this problem from a top-down perspective. Using well-motivated, yet theoretically broadly defined assumptions about the UV theory, we analyse their effective dimension-six interactions via an one-loop matching procedure. We find that the $Q_{\widetilde{W}}$ is either absent, or at least two-loop suppressed for scenarios: Heavy scalar extensions of the SM do not generate triple field strength and gauge-Higgs CPV SMEFT dimension-six operators. On the other hand, one needs to extend the SM by at least two heavy vector-like fermions to generate these operators at two-loop-level. The non-SM CPV Yukawa couplings play a crucial role to induce the rank-four Levi-Civita tensor, and thus the dual of the field strength tensors.

The phenomenological consequence of this is that new CP-violating electroweak physics can in principle be captured via a combination of Higgs and diboson analyses. The diboson sensitivity at the LHC is driven by the large cross section and the relatively clean $W\gamma$ channels, which serve to predominantly constrain the $Q_{H\widetilde{W}B}$, which is related to a blind direction of Higgs data. The possibility to omit contributions $Q_{\widetilde{W}}$ is therefore critical and allows to constrain the remaining operators at dimension-six, one-loop matched level, {\it i.e.} at the theoretical level where the BSM modifications are expected to be sizable.

\acknowledgments
We would like to thank Andy Buckley for helpful discussions. 
The work of S.D.B. and J.C. is supported by the Science and Engineering Research Board, Government of India, under the agreements SERB/PHY/2016348 (Early Career Research Award) and SERB/PHY/2019501 (MATRICS).
C.E. is supported by the UK Science Technologies and Facilities Council (STFC) under grant ST/T000945/1 and by the IPPP Associateship Scheme. 
M.S. is supported by the STFC under grant ST/P001246/1. 
P.S. is supported by an STFC studentship under grant ST/T506102/1.

\appendix
\section{$ Q_{\widetilde{W}} $ operator at two-loop processes}\label{sec:two-loop-calc}
\begin{figure}[!t]
	\includegraphics[height=10cm,width=15cm]{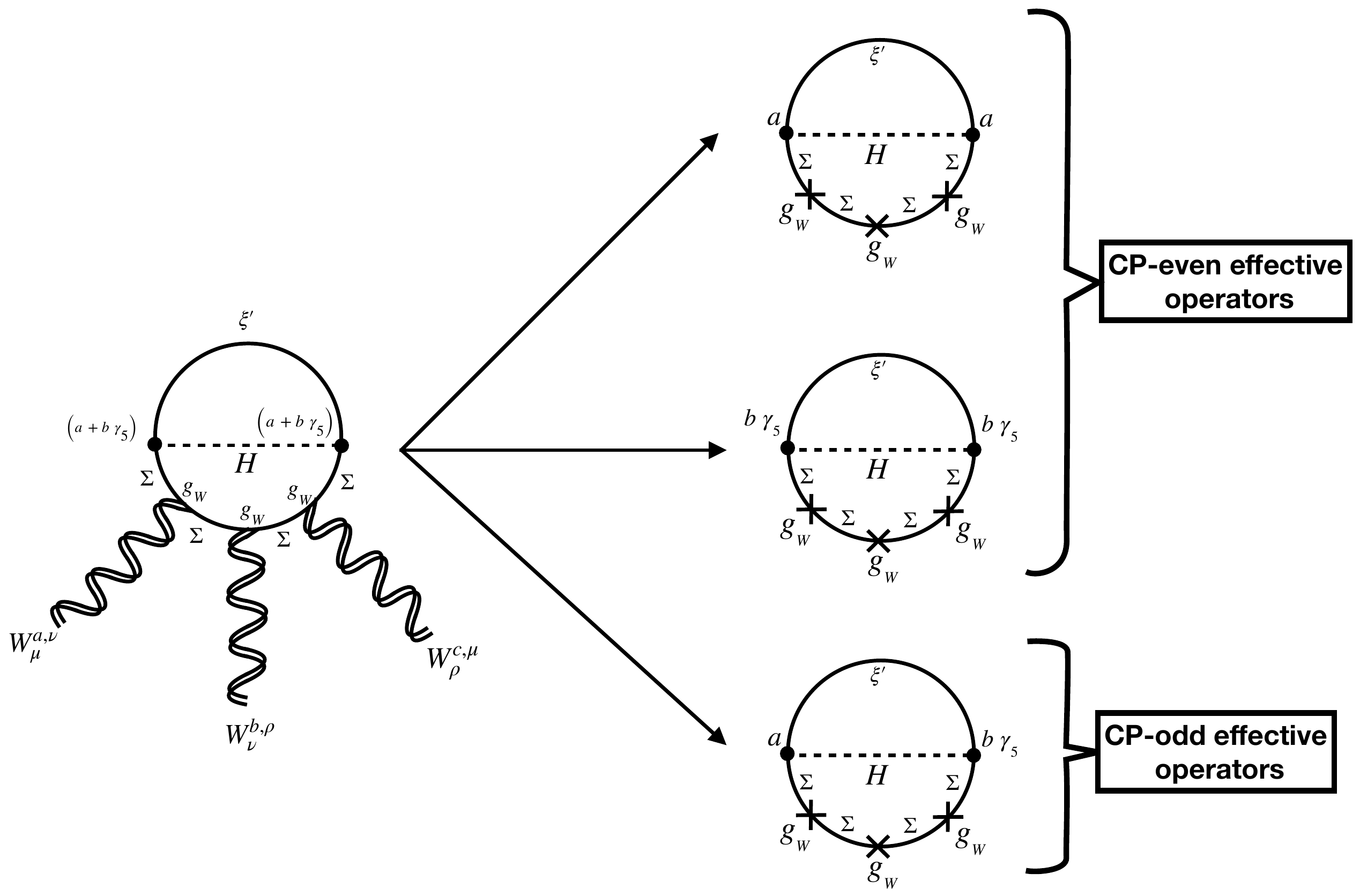}
	\caption{Two-loop diagram generating the SMEFT Warsaw basis $ Q_W $ and $ Q_{\widetilde{W}} $ operators by integrating out heavy Doublet-Singlet VLL, see Eq.~\eqref{eq:vlike}. Note that only the heavy isospin-doublet lepton couples to the $ SU(2)_L $ field strength tensors. The Yukawa vertices are shown in generic manner in this figure. The `$ a $' and `$ b $' are defined according to the Yukawa terms in the Lagrangian.}
	\label{fig:qwtil-VLL}
\end{figure}

%
%
In this section, we discuss the origin of the CPV $ X^3 $ operators that are absent at the one-loop matching of different BSMs to SMEFT~\cite{Henning:2014wua,Bakshi:2020eyg,Anisha:2020ggj,Bakshi:2018ics,Gherardi:2020det}.
 We have considered the heavy fermion extensions of the SM in Sec.~\ref{sec:theory}, and have argued that these CPV operators are generated  in the process of integrating out, and matching to SMEFT only at two-loop and beyond. We further note that these operators cannot be generated by integrating out only heavy scalars (up to two-loop at least) due to the lack of presence of $\gamma_{_5}$ and, therefore, the fourth-ranked antisymmetric tensor to form the dual field strength. Thus, CP-violation in the scalar potential cannot be captured through these operators. The presence of this operator, certainly, signifies the CP-violation in the non-SM Yukawa interactions.  In the subsequent discussion, we focus on the specific  VLL models \cite{DasBakshi:2020ejz,Joglekar:2012vc,Angelescu:2018dkk,Chala:2020odv,Bissmann:2020lge}, and outline the dominant emergence of $ Q_{\widetilde{W}} $ from two-loop processes for the first time. Note that in an EFT context, the `pinching' the Higgs propagator in Fig.~\ref{fig:qwtil-VLL} indicates the emergence of the considered operator under RGE flow. While this is a technically challenging subject in its own right in EFT discussion, such issues are absent when we consider UV-complete scenarios, see also~\cite{Englert:2019rga}.

\subsection{Doublet-Singlet VLL}
We work with a BSM scenario where the SM is extended by one isospin-doublet and two isospin-singlet heavy VLL having the following SM gauge quantum number:
\begin{equation}
\Sigma_{L,R}=\begin{pmatrix} \eta \\ \xi	\end{pmatrix}_{L,R} : (1,2,\mathcal{Y}), \;\;
\eta'_{L,R} : (1,1,\mathcal{Y}+\frac{1}{2}),\;\;
\xi'_{L,R} : (1,1,\mathcal{Y}-\frac{1}{2}).
\end{equation}
The relevant part of the BSM Lagrangian involving these heavy leptons is given by \cite{DasBakshi:2020ejz}
\begin{multline}\label{eq:vlike}
	\mathcal{L}_{\text{DS}}  =  \bar{\Sigma} ( i \D_{_\Sigma} - m_{_{\Sigma}}) \Sigma + \bar{\eta'} ( i \D_{\eta} - m_{{\eta}}) \eta' + \bar{\xi'} ( i \D_{\xi} - m_{{\xi}}) \xi' \\
	- \left\lbrace \bar{\Sigma} \tilde{H} ( Y_{\eta_{_L}} \mathbb{P}_L + Y_{\eta_{_R}} \mathbb{P}_{R}) \eta' + \bar{\Sigma} H ( Y_{\xi_{_L}} \mathbb{P}_{L} + Y_{\xi_{_R}} \mathbb{P}_{R}) \xi' + \text{h.c.} \right\rbrace.
\end{multline}

\begin{figure}[!b]
	\centering
	\includegraphics[width=0.8\linewidth]{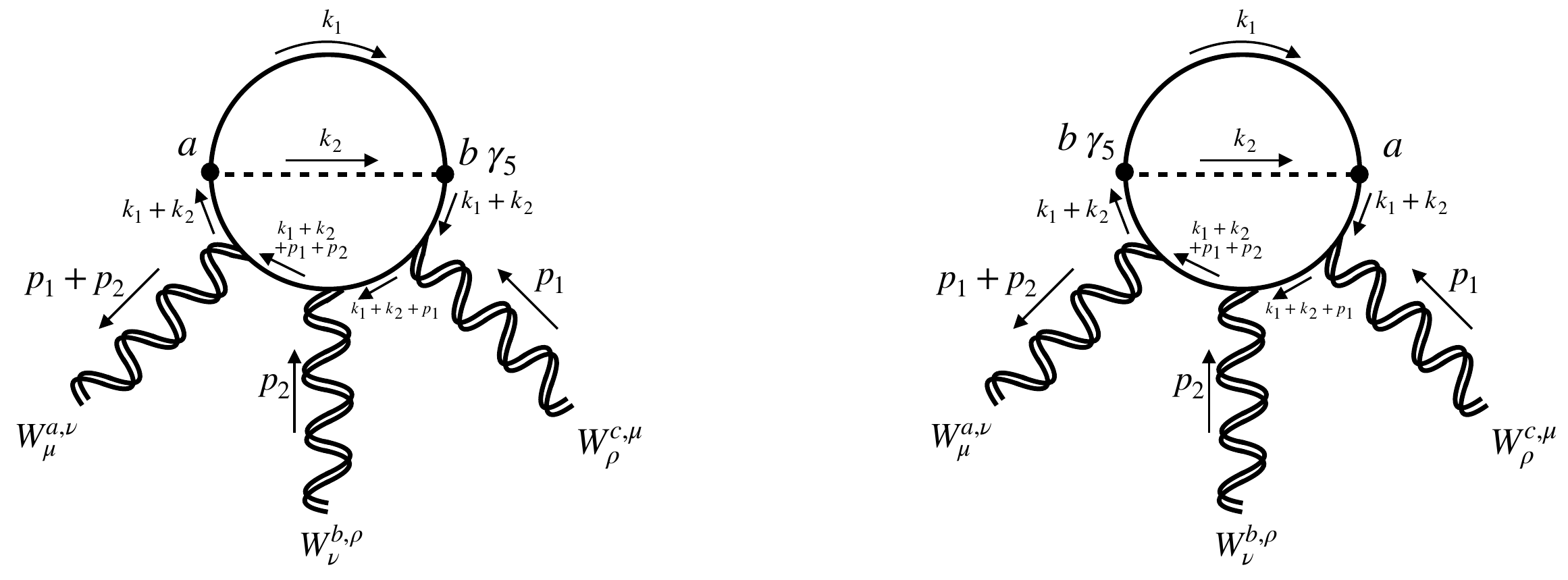}
	\caption{Possible configurations of the two-loop diagram contributing to the $ Q_{\widetilde{W}} $ operator. The internal (loop) and external momentums are explicitly shown.}
	\label{fig:cwtilde-vll-mom}
\end{figure}

Here, we dedicate our analysis to demonstrate the emergence of  $ Q_{\widetilde{W}}\,  (=\epsilon^{IJK} \widetilde{W}^{I\nu}_{\mu} W^{J\rho}_{\nu} W^{K\mu}_{\rho})$ at two-loop level, where $ \widetilde{W}^{I\nu}_{\mu}$ represents the dual of the field strength tensor $ W^{I\nu}_{\mu} $. The origin of $ Q_{\widetilde{W}} $ operator is depicted in Fig.~\ref{fig:qwtil-VLL}. We depict the Yukawa vertices in the two-loop diagram in terms of `$ a $' and `$ b $'  instead of parameters of any specific model to note down the results in more  generic form and readily applicable to similar models of different couplings.\footnote{We expand the Yukawa vertex and collect the coefficients of $ \mathbb{1} $ and $ \gamma_5 $ to determine `$ a $' and `$ b $' respectively. This convention is introduced and explained in Eq.~\eqref{eq:twoheavyfer}.} The contribution of this two-loop diagram can be captured through the following integral\footnote{Here, we use an identity involving $\gamma_{_5}$ in (3+1)-dimension to rewrite $ \sigma_{\mu\nu}  \gamma_5 $ in terms of the fourth rank Levi-Civita tensor before proceeding to the trace calculation \cite{Jegerlehner:2000dz,Chanowitz:1979zu,Ahmed:2020kme,Heller:2020owb}. Then, one can continue this integration to the $ D $ dimension following t’ Hooft-Veltman dimensional regularization method.},
\begin{align}
\mathcal{I}(p_1,p_2;m_{\xi},m_{_\Sigma})=\int\frac{d^4 k_1}{(2\pi)^4}\frac{d^4 k_2}{(2\pi)^4} \text{Tr}&\left[\frac{1}{\slashed{k}_1-m_{\xi}}\left(a + b \gamma_5\right)\frac{1}{k_2^2-m^2_{_H}}\frac{1}{\slashed{k}_1+\slashed{k}_2-m_{_\Sigma}}\left(\sigma_{\mu\nu} W^{I\mu\nu}\tau^I\right)\right.\nonumber\\
&\left. \frac{1}{\slashed{k}_1+\slashed{k}_2+\slashed{p}_1-m_{_\Sigma}} \left(\sigma_{\rho\sigma} W^{J\rho\sigma}\tau^J\right)\frac{1}{\slashed{k}_1+\slashed{k}_2+\slashed{p}_1+\slashed{p}_2-m_{_\Sigma}}\right.\nonumber\\
&\left.\left(\sigma_{\alpha\beta} W^{K\alpha\beta}\tau^K\right)\frac{1}{\slashed{k}_1+\slashed{k}_2-m_{_\Sigma}}\left(a +b\gamma_5\right)\right],
\end{align}
where, $ \tau^i $s are normalised $SU(2)$ generators ($=\sigma^i/2 $). The momentum configurations are shown in Fig.~\ref{fig:cwtilde-vll-mom}. Then, we collect the terms leading to the CPV effective operator $ Q_{\widetilde{W}} $ through two following integrals:
\begin{align}
\mathcal{I}_1^{\cancel{\text{CP}}}(p_1,p_2;m_{\xi},m_{_\Sigma})=\int\frac{d^4 k_1}{(2\pi)^4}\frac{d^4 k_2}{(2\pi)^4} \text{Tr}&\left[\frac{1}{\slashed{k}_1-m_{\xi}}a \frac{1}{k_2^2-m^2_{_H}}\frac{1}{\slashed{k}_1+\slashed{k}_2-m_{_\Sigma}}\left(\sigma_{\mu\nu} W^{I\mu\nu}\tau^I\right)\right.\nonumber\\
&\left. \frac{1}{\slashed{k}_1+\slashed{k}_2+\slashed{p}_1-m_{_\Sigma}} \left(\sigma_{\rho\sigma} W^{J\rho\sigma}\tau^J\right)\frac{1}{\slashed{k}_1+\slashed{k}_2+\slashed{p}_1+\slashed{p}_2-m_{_\Sigma}}\right.\nonumber\\
&\left.\left(\sigma_{\alpha\beta} W^{K\alpha\beta}\tau^K\right)\frac{1}{\slashed{k}_1+\slashed{k}_2-m_{_\Sigma}}b\gamma_5\right],
\end{align}
and,
\begin{align}
\mathcal{I}_2^{\cancel{\text{CP}}}(p_1,p_2;m_{\xi},m_{_\Sigma})=\int\frac{d^4 k_1}{(2\pi)^4}\frac{d^4 k_2}{(2\pi)^4} \text{Tr}&\left[\frac{1}{\slashed{k}_1-m_{\xi}} b \gamma_5 \frac{1}{k_2^2-m^2_{_H}}\frac{1}{\slashed{k}_1+\slashed{k}_2-m_{_\Sigma}}\left(\sigma_{\mu\nu} W^{I\mu\nu}\tau^I\right)\right.\nonumber\\
&\left. \frac{1}{\slashed{k}_1+\slashed{k}_2+\slashed{p}_1-m_{_\Sigma}} \left(\sigma_{\rho\sigma} W^{J\rho\sigma}\tau^J\right)\frac{1}{\slashed{k}_1+\slashed{k}_2+\slashed{p}_1+\slashed{p}_2-m_{_\Sigma}}\right.\nonumber\\
&\left.\left(\sigma_{\alpha\beta} W^{K\alpha\beta}\tau^K\right)\frac{1}{\slashed{k}_1+\slashed{k}_2-m_{_\Sigma}}a\right].
\end{align}

\allowdisplaybreaks

Now, we proceed to solve $ \mathcal{I}_1^{\cancel{\text{CP}}} $ further using the identities, $ \sigma_{\mu\nu} \gamma_5 = -\frac{i}{2} \levifo \sigma^{\rho\sigma} $ and $ \text{Tr}\left[\tau^I\tau^J\tau^K\right] = \frac{i}{4} \epsilon^{IJK}$, and we rewrite,
\begin{align}\label{eq:int-I1}
\mathcal{I}_1^{\cancel{\text{CP}}}(p_1,p_2;m_{\xi},m_{_\Sigma})=& \frac{i}{4} a\, b\, \epsilon^{IJK} W^{I\mu\nu} W^{J\rho\sigma} W^{K\alpha\beta} \int\frac{d^4 k_1}{(2\pi)^4}\frac{d^4 k_2}{(2\pi)^4} \frac{1}{k_2^2-m^2_{_H}} \text{Tr}\left[\frac{1}{\slashed{k}_1-m_{\xi}} \frac{1}{\slashed{k}_1+\slashed{k}_2-m_{_\Sigma}}\sigma_{\mu\nu}\right.\nonumber\\
&\left.  \frac{1}{\slashed{k}_1+\slashed{k}_2+\slashed{p}_1-m_{_\Sigma}} \sigma_{\rho\sigma}\frac{1}{\slashed{k}_1+\slashed{k}_2+\slashed{p}_1+\slashed{p}_2-m_{_\Sigma}}\sigma_{\alpha\beta}\gamma_5\frac{1}{-\slashed{k}_1-\slashed{k}_2-m_{_\Sigma}}\right]\nonumber\\
=&\frac{a\, b\,}{8}  \epsilon^{IJK} W^{I\mu\nu} W^{J\rho\sigma} \widetilde{W}^{K\alpha\beta} \int\frac{d^4 k_1}{(2\pi)^4}\frac{d^4 k_2}{(2\pi)^4} \frac{1}{k_2^2-m^2_{_H}} \text{Tr}\left[\frac{1}{\slashed{k}_1-m_{\xi}} \frac{1}{\slashed{k}_1+\slashed{k}_2-m_{_\Sigma}}\sigma_{\mu\nu}\right.\nonumber\\
&\left.  \frac{1}{\slashed{k}_1+\slashed{k}_2+\slashed{p}_1-m_{_\Sigma}} \sigma_{\rho\sigma}\frac{1}{\slashed{k}_1+\slashed{k}_2+\slashed{p}_1+\slashed{p}_2-m_{_\Sigma}}\sigma_{\alpha\beta}\frac{1}{-\slashed{k}_1-\slashed{k}_2-m_{_\Sigma}}\right]\nonumber\\
=&-\frac{a\, b\,}{8} \epsilon^{IJK} W^{I\mu\nu} W^{J\rho\sigma} \widetilde{W}^{K\alpha\beta} \int\frac{d^4 k_1}{(2\pi)^4}\frac{d^4 k_2}{(2\pi)^4} \frac{1}{k_2^2-m^2_{_H}}\frac{\text{Tr}\bigg[\left(\slashed{k}_1+m_{\xi}\right)\left(\slashed{k}_1+\slashed{k}_2+m_{_\Sigma}\right)}{\left(k_1^2-m_\xi^2\right)\left(\left(k_1+k_2\right)^2-m_{_\Sigma}^2\right)}\nonumber\\ &\frac{\sigma_{\mu\nu}\left(\slashed{k}_1+\slashed{k}_1+\slashed{p}_1+m_{_\Sigma}\right)\sigma_{\rho\sigma}\left(\slashed{k}_1+\slashed{k}_2+\slashed{p}_1+\slashed{p}_2+m_{_\Sigma}\right)\sigma_{\alpha\beta}\left(\slashed{k}_1+\slashed{k}_2-m_{_\Sigma}\right)\bigg]}{\left(\left(k_1+k_2+p_1\right)^2-m_{_\Sigma}^2\right)\left(\left(k_1+k_2+p_1+p_2\right)^2-m_{_\Sigma}^2\right)\left(\left(k_1+k_2\right)^2-m_{_\Sigma}^2\right)}
\end{align}
\begin{align}\label{eq:appdsvlli1}
\supset&-\frac{a\, b\,}{8} \epsilon^{IJK} W^{I\mu\nu} W^{J\rho\sigma} \widetilde{W}^{K\alpha\beta} \int\frac{d^4 k_1}{(2\pi)^4}\frac{d^4 k_2}{(2\pi)^4} \frac{1}{k_2^2-m^2_{_H}}\frac{1}{\left(k_1^2-m_\xi^2\right)\left(\left(k_1+k_2\right)^2-m_{_\Sigma}^2\right)}\nonumber\\ &\frac{\left(4\, i\, g_{\sigma\alpha}\, g_{\beta\mu}\, g_{\nu\rho}\,\right)\left[ m_{\xi}^3 m_{\Sigma}^3 - m_{\xi}^2\, m_{\Sigma} k_1^2\, + m_{\Sigma}^3 k_1^2 + \ldots \right]}{\left(\left(k_1+k_2+p_1\right)^2-m_{_\Sigma}^2\right)\left(\left(k_1+k_2+p_1+p_2\right)^2-m_{_\Sigma}^2\right)\left(\left(k_1+k_2\right)^2-m_{_\Sigma}^2\right)}\nonumber\\
=&-\frac{i\, a\, b\,}{2} \epsilon^{IJK} W^{I\mu}{}_{\nu} W^{J\nu}{}_{\rho} \widetilde{W}^{K\rho}{}_{\mu} \int\frac{d^4 k_1}{(2\pi)^4}\frac{d^4 k_2}{(2\pi)^4} \frac{1}{k_2^2-m^2_{_H}}\frac{1}{\left(k_1^2-m_\xi^2\right)\left(\left(k_1+k_2\right)^2-m_{_\Sigma}^2\right)}\nonumber\\ &\frac{\left[ m_{\xi}^3 m_{\Sigma}^3 - m_{\xi}^2\, m_{\Sigma} k_1^2\, + m_{\Sigma}^3 k_1^2 + \ldots \right]}{\left(\left(k_1+k_2+p_1\right)^2-m_{_\Sigma}^2\right)\left(\left(k_1+k_2+p_1+p_2\right)^2-m_{_\Sigma}^2\right)\left(\left(k_1+k_2\right)^2-m_{_\Sigma}^2\right)}.
\end{align}
Here, instead of computing the full momentum integral, we have highlighted the part that leads to the $ Q_{\widetilde{W}} $ operator.  We have used Package-X~\cite{Patel:2016fam} to cross-check our results. We perform similar calculations for $ \mathcal{I}_2^{\cancel{\text{CP}}} $,
\begin{align}
\mathcal{I}_2^{\cancel{\text{CP}}}(p_1,p_2;m_{\xi},m_{_\Sigma})=& \frac{i}{4} a\, b\, \epsilon^{IJK} W^{I\mu\nu} W^{J\rho\sigma} W^{K\alpha\beta} \int\frac{d^4 k_1}{(2\pi)^4}\frac{d^4 k_2}{(2\pi)^4} \frac{1}{k_2^2-m^2_{_H}} \text{Tr}\left[\frac{1}{\slashed{k}_1-m_{\xi}} \gamma_5\frac{1}{\slashed{k}_1+\slashed{k}_2-m_{_\Sigma}}\sigma_{\mu\nu}\right.\nonumber\\
&\left.  \frac{1}{\slashed{k}_1+\slashed{k}_2+\slashed{p}_1-m_{_\Sigma}} \sigma_{\rho\sigma}\frac{1}{\slashed{k}_1+\slashed{k}_2+\slashed{p}_1+\slashed{p}_2-m_{_\Sigma}}\sigma_{\alpha\beta}\frac{1}{\slashed{k}_1+\slashed{k}_2-m_{_\Sigma}}\right]\nonumber\\
=&\frac{a\, b\,}{8}  \epsilon^{IJK} W^{I\mu\nu} W^{J\rho\sigma} \widetilde{W}^{K\alpha\beta} \int\frac{d^4 k_1}{(2\pi)^4}\frac{d^4 k_2}{(2\pi)^4} \frac{1}{k_2^2-m^2_{_H}} \text{Tr}\left[\frac{1}{\slashed{k}_1-m_{\xi}} \frac{1}{-\slashed{k}_1-\slashed{k}_2-m_{_\Sigma}}\gamma_5\sigma_{\mu\nu}\right.\nonumber\\
&\left.  \frac{1}{\slashed{k}_1+\slashed{k}_2+\slashed{p}_1-m_{_\Sigma}} \sigma_{\rho\sigma}\frac{1}{\slashed{k}_1+\slashed{k}_2+\slashed{p}_1+\slashed{p}_2-m_{_\Sigma}}\sigma_{\alpha\beta}\frac{1}{\slashed{k}_1+\slashed{k}_2-m_{_\Sigma}}\right]\nonumber\\
=&-\frac{a\, b\,}{8} \epsilon^{IJK} W^{I\mu\nu} W^{J\rho\sigma} \widetilde{W}^{K\alpha\beta} \int\frac{d^4 k_1}{(2\pi)^4}\frac{d^4 k_2}{(2\pi)^4} \frac{1}{k_2^2-m^2_{_H}}\frac{\text{Tr}\bigg[\left(\slashed{k}_1+m_{\xi}\right)\left(\slashed{k}_1+\slashed{k}_2-m_{_\Sigma}\right)}{\left(k_1^2-m_\xi^2\right)\left(\left(k_1+k_2\right)^2-m_{_\Sigma}^2\right)}\nonumber\\ &\frac{\sigma_{\mu\nu}\left(\slashed{k}_1+\slashed{k}_1+\slashed{p}_1+m_{_\Sigma}\right)\sigma_{\rho\sigma}\left(\slashed{k}_1+\slashed{k}_2+\slashed{p}_1+\slashed{p}_2+m_{_\Sigma}\right)\sigma_{\alpha\beta}\left(\slashed{k}_1+\slashed{k}_2+m_{_\Sigma}\right)\bigg]}{\left(\left(k_1+k_2+p_1\right)^2-m_{_\Sigma}^2\right)\left(\left(k_1+k_2+p_1+p_2\right)^2-m_{_\Sigma}^2\right)\left(\left(k_1+k_2\right)^2-m_{_\Sigma}^2\right)}
\end{align}
\begin{align}\label{eq:appdsvlli2}
\supset&-\frac{a\, b\,}{8} \epsilon^{IJK} W^{I\mu\nu} W^{J\rho\sigma} \widetilde{W}^{K\alpha\beta} \int\frac{d^4 k_1}{(2\pi)^4}\frac{d^4 k_2}{(2\pi)^4} \frac{1}{k_2^2-m^2_{_H}}\frac{1}{\left(k_1^2-m_\xi^2\right)\left(\left(k_1+k_2\right)^2-m_{_\Sigma}^2\right)}\nonumber\\ &\frac{\left(4\, i\, g_{\sigma\alpha}\, g_{\beta\mu}\, g_{\nu\rho}\,\right)\left[ m_{\xi}^3 m_{\Sigma}^3 + 3 m_{\xi}^2\, m_{\Sigma} k_1^2\, - m_{\Sigma}^3 k_1^2 + \ldots \right]}{\left(\left(k_1+k_2+p_1\right)^2-m_{_\Sigma}^2\right)\left(\left(k_1+k_2+p_1+p_2\right)^2-m_{_\Sigma}^2\right)\left(\left(k_1+k_2\right)^2-m_{_\Sigma}^2\right)}\nonumber\\
=&-\frac{i\, a\, b\,}{2} \epsilon^{IJK} W^{I\mu}{}_{\nu} W^{J\nu}{}_{\rho} \widetilde{W}^{K\rho}{}_{\mu} \int\frac{d^4 k_1}{(2\pi)^4}\frac{d^4 k_2}{(2\pi)^4} \frac{1}{k_2^2-m^2_{_H}}\frac{1}{\left(k_1^2-m_\xi^2\right)\left(\left(k_1+k_2\right)^2-m_{_\Sigma}^2\right)}\nonumber\\
 &\frac{\left[ m_{\xi}^3 m_{\Sigma}^3 + 3 m_{\xi}^2\, m_{\Sigma} k_1^2\, - m_{\Sigma}^3 k_1^2 + \ldots \right]}{\left(\left(k_1+k_2+p_1\right)^2-m_{_\Sigma}^2\right)\left(\left(k_1+k_2+p_1+p_2\right)^2-m_{_\Sigma}^2\right)\left(\left(k_1+k_2\right)^2-m_{_\Sigma}^2\right)}.
\end{align}

\

Here, we choose not to show the contribution from the $\Sigma - \eta'$ two-loop diagram separately as $\Sigma - \eta'$ contribution can be derived by replacing $ m_{\xi} \rightarrow m_{\eta}$ in Eqs.~\eqref{eq:appdsvlli1} and \eqref{eq:appdsvlli2}.

\

\subsection{Doublet-Triplet VLL}
\begin{figure}
	\centering
	\includegraphics[width=0.8\linewidth]{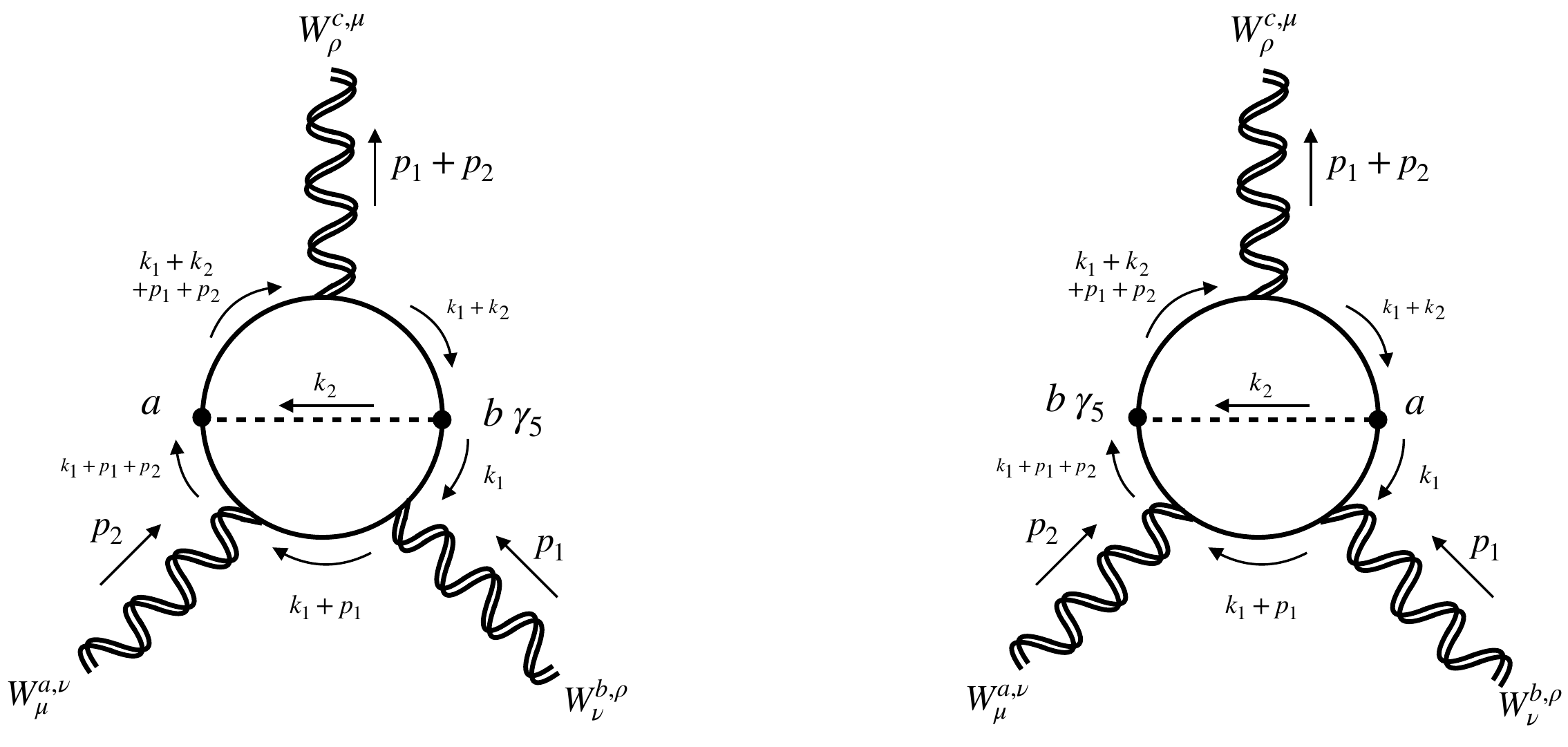}
	\caption{Diagrams showing the momentum configurations for the two-loop diagram in the Doublet-Triplet VLL model.}
	\label{fig:cwtilde-dtvll-mom}
\end{figure}
We present the contribution from the two-loop diagram produced in the particular VLL model discussed in Sec.~\ref{sec:theory}. The corresponding two-loop diagram is discussed in Fig.~\ref{fig:twoloopDT}, and the momentum configurations are available in Fig.~\ref{fig:cwtilde-dtvll-mom}. Similar to the earlier case, the relevant loop integral is given by
\begin{align}
&\mathcal{I}_{{\text{DT}}}(p_1,p_2;m_{\chi_{_2}},m_{\chi_{_3}})=\int\frac{d^4 k_1}{(2\pi)^4}\frac{d^4 k_2}{(2\pi)^4} \text{Tr}\left[\frac{1}{k_2^2-m^2_{_H}} \frac{1}{\slashed{k}_1-m_{\chi_{_3}}}\left(\sigma_{\mu\nu} W^{I\mu\nu}\tau^I\right)\frac{1}{\slashed{k}_1+\slashed{p}_1-m_{\chi_{_3}}} \left(\sigma_{\rho\sigma} W^{J\rho\sigma}\tau^J\right) \right.\nonumber\\
&\left. \frac{1}{\slashed{k}_1+\slashed{p}_1+\slashed{p}_2-m_{\chi_{_3}}} \left(a + b \gamma_5\right) \frac{1}{\slashed{k}_1+\slashed{k}_2+\slashed{p}_1+\slashed{p}_2-m_{\chi_{_2}}}\left(\sigma_{\alpha\beta} W^{K\alpha\beta}\tau^K\right)\frac{1}{\slashed{k}_1+\slashed{k}_2-m_{\chi_{_2}}}\left(a +b\gamma_5\right)\right]
\end{align}
\begin{align}
\supset& - i\, a\, b\, \epsilon^{IJK} W^{I\mu}{}_{\nu} W^{J\nu}{}_{\rho} \widetilde{W}^{K\rho}{}_{\mu} \int\frac{d^4 k_1}{(2\pi)^4}\frac{d^4 k_2}{(2\pi)^4} \frac{1}{k_2^2-m^2_{_H}} \frac{1}{\left(k_1^2-m_{\chi_{_3}}^2\right)\left(\left(k_1+p_1\right)^2-m_{\chi_{_3}}^2\right)}\nonumber\\ 
&\frac{\left[m_{\chi_{_2}}^3 m_{\chi_{_3}}^2 + \ldots\right]}{\left(\left(k_1+p_1+p_2\right)^2-m_{\chi_{_3}}^2\right)\left(\left(k_1+k_2+p_1+p_2\right)^2-m_{\chi_{_2}}^2\right)\left(\left(k_1+k_2\right)^2-m_{\chi_{_2}}^2\right)} \ .
\end{align}
We perform the derivation similar to the previous subsection and deduce the above equation. Unlike the Doublet-Singlet VLL model, in this model the $ SU(2)_L $ field strength tensor can couple to both the leptons in the loop. Therefore, we get a larger number of possible two-loop diagrams in this model in comparison to the Doublet-Singlet model, where the field strength tensor couples only to the isospin-doublet lepton.
Also following similar arguments, one can conclude that the CPV triple gluon field strength operator $Q_{\widetilde{G}} \ ( = f^{ABC} \widetilde{G}^{A\nu}_{\mu} G^{B\rho}_{\nu} G^{C\mu}_{\rho} )$ is generated in BSMs with heavy colored fermions.

\bibliography{paper.bbl}

\begin{thebibliography}{99}
\expandafter\ifx\csname natexlab\endcsname\relax\def\natexlab#1{#1}\fi
\expandafter\ifx\csname bibnamefont\endcsname\relax
  \def\bibnamefont#1{#1}\fi
\expandafter\ifx\csname bibfnamefont\endcsname\relax
  \def\bibfnamefont#1{#1}\fi
\expandafter\ifx\csname citenamefont\endcsname\relax
  \def\citenamefont#1{#1}\fi
\expandafter\ifx\csname url\endcsname\relax
  \def\url#1{\texttt{#1}}\fi
\expandafter\ifx\csname urlprefix\endcsname\relax\def\urlprefix{URL }\fi
\providecommand{\bibinfo}[2]{#2}
\providecommand{\eprint}[2][]{\url{#2}}

\bibitem[{\citenamefont{Weinberg}(1979)}]{Weinberg:1978kz}
\bibinfo{author}{\bibfnamefont{S.}~\bibnamefont{Weinberg}},
  \bibinfo{journal}{Physica} \textbf{\bibinfo{volume}{A96}},
  \bibinfo{pages}{327} (\bibinfo{year}{1979}).

\bibitem[{\citenamefont{Brivio and Trott}(2019)}]{Brivio:2017vri}
\bibinfo{author}{\bibfnamefont{I.}~\bibnamefont{Brivio}} \bibnamefont{and}
  \bibinfo{author}{\bibfnamefont{M.}~\bibnamefont{Trott}},
  \bibinfo{journal}{Phys. Rept.} \textbf{\bibinfo{volume}{793}},
  \bibinfo{pages}{1} (\bibinfo{year}{2019}), \eprint{1706.08945}.

\bibitem[{\citenamefont{Dawson et~al.}(2019)\citenamefont{Dawson, Englert, and
  Plehn}}]{Dawson:2018dcd}
\bibinfo{author}{\bibfnamefont{S.}~\bibnamefont{Dawson}},
  \bibinfo{author}{\bibfnamefont{C.}~\bibnamefont{Englert}}, \bibnamefont{and}
  \bibinfo{author}{\bibfnamefont{T.}~\bibnamefont{Plehn}},
  \bibinfo{journal}{Phys. Rept.} \textbf{\bibinfo{volume}{816}},
  \bibinfo{pages}{1} (\bibinfo{year}{2019}), \eprint{1808.01324}.

\bibitem[{\citenamefont{Aad et~al.}(2020{\natexlab{a}})}]{Aad:2020sle}
\bibinfo{author}{\bibfnamefont{G.}~\bibnamefont{Aad}} \bibnamefont{et~al.}
  (\bibinfo{collaboration}{ATLAS}) (\bibinfo{year}{2020}{\natexlab{a}}),
  \eprint{2006.15458}.

\bibitem[{\citenamefont{Lehman and Martin}(2016)}]{Lehman:2015coa}
\bibinfo{author}{\bibfnamefont{L.}~\bibnamefont{Lehman}} \bibnamefont{and}
  \bibinfo{author}{\bibfnamefont{A.}~\bibnamefont{Martin}},
  \bibinfo{journal}{JHEP} \textbf{\bibinfo{volume}{02}}, \bibinfo{pages}{081}
  (\bibinfo{year}{2016}), \eprint{1510.00372}.

\bibitem[{\citenamefont{Murphy}(2020{\natexlab{a}})}]{Murphy:2020cly}
\bibinfo{author}{\bibfnamefont{C.~W.} \bibnamefont{Murphy}}
  (\bibinfo{year}{2020}{\natexlab{a}}), \eprint{2012.13291}.

\bibitem[{\citenamefont{Murphy}(2020{\natexlab{b}})}]{Murphy:2020rsh}
\bibinfo{author}{\bibfnamefont{C.~W.} \bibnamefont{Murphy}},
  \bibinfo{journal}{JHEP} \textbf{\bibinfo{volume}{10}}, \bibinfo{pages}{174}
  (\bibinfo{year}{2020}{\natexlab{b}}), \eprint{2005.00059}.

\bibitem[{\citenamefont{Li et~al.}(2020)\citenamefont{Li, Ren, Shu, Xiao, Yu,
  and Zheng}}]{Li:2020gnx}
\bibinfo{author}{\bibfnamefont{H.-L.} \bibnamefont{Li}},
  \bibinfo{author}{\bibfnamefont{Z.}~\bibnamefont{Ren}},
  \bibinfo{author}{\bibfnamefont{J.}~\bibnamefont{Shu}},
  \bibinfo{author}{\bibfnamefont{M.-L.} \bibnamefont{Xiao}},
  \bibinfo{author}{\bibfnamefont{J.-H.} \bibnamefont{Yu}}, \bibnamefont{and}
  \bibinfo{author}{\bibfnamefont{Y.-H.} \bibnamefont{Zheng}}
  (\bibinfo{year}{2020}), \eprint{2005.00008}.

\bibitem[{\citenamefont{Biekötter et~al.}(2021)\citenamefont{Biekötter,
  Gregg, Krauss, and Schönherr}}]{Biekotter:2021int}
\bibinfo{author}{\bibfnamefont{A.}~\bibnamefont{Biekötter}},
  \bibinfo{author}{\bibfnamefont{P.}~\bibnamefont{Gregg}},
  \bibinfo{author}{\bibfnamefont{F.}~\bibnamefont{Krauss}}, \bibnamefont{and}
  \bibinfo{author}{\bibfnamefont{M.}~\bibnamefont{Schönherr}}
  (\bibinfo{year}{2021}), \eprint{2102.01115}.

\bibitem[{\citenamefont{Corbett et~al.}(2021)\citenamefont{Corbett, Helset,
  Martin, and Trott}}]{Corbett:2021eux}
\bibinfo{author}{\bibfnamefont{T.}~\bibnamefont{Corbett}},
  \bibinfo{author}{\bibfnamefont{A.}~\bibnamefont{Helset}},
  \bibinfo{author}{\bibfnamefont{A.}~\bibnamefont{Martin}}, \bibnamefont{and}
  \bibinfo{author}{\bibfnamefont{M.}~\bibnamefont{Trott}}
  (\bibinfo{year}{2021}), \eprint{2102.02819}.

\bibitem[{\citenamefont{Burges and Schnitzer}(1983)}]{Burges:1983zg}
\bibinfo{author}{\bibfnamefont{C.~J.~C.} \bibnamefont{Burges}}
  \bibnamefont{and} \bibinfo{author}{\bibfnamefont{H.~J.}
  \bibnamefont{Schnitzer}}, \bibinfo{journal}{Nucl. Phys.}
  \textbf{\bibinfo{volume}{B228}}, \bibinfo{pages}{464} (\bibinfo{year}{1983}).

\bibitem[{\citenamefont{Leung et~al.}(1986)\citenamefont{Leung, Love, and
  Rao}}]{Leung:1984ni}
\bibinfo{author}{\bibfnamefont{C.~N.} \bibnamefont{Leung}},
  \bibinfo{author}{\bibfnamefont{S.~T.} \bibnamefont{Love}}, \bibnamefont{and}
  \bibinfo{author}{\bibfnamefont{S.}~\bibnamefont{Rao}}, \bibinfo{journal}{Z.
  Phys.} \textbf{\bibinfo{volume}{C31}}, \bibinfo{pages}{433}
  (\bibinfo{year}{1986}).

\bibitem[{\citenamefont{Buchmuller and Wyler}(1986)}]{Buchmuller:1985jz}
\bibinfo{author}{\bibfnamefont{W.}~\bibnamefont{Buchmuller}} \bibnamefont{and}
  \bibinfo{author}{\bibfnamefont{D.}~\bibnamefont{Wyler}},
  \bibinfo{journal}{Nucl. Phys. B} \textbf{\bibinfo{volume}{268}},
  \bibinfo{pages}{621} (\bibinfo{year}{1986}).

\bibitem[{\citenamefont{Hagiwara et~al.}(1987)\citenamefont{Hagiwara, Peccei,
  Zeppenfeld, and Hikasa}}]{Hagiwara:1986vm}
\bibinfo{author}{\bibfnamefont{K.}~\bibnamefont{Hagiwara}},
  \bibinfo{author}{\bibfnamefont{R.~D.} \bibnamefont{Peccei}},
  \bibinfo{author}{\bibfnamefont{D.}~\bibnamefont{Zeppenfeld}},
  \bibnamefont{and} \bibinfo{author}{\bibfnamefont{K.}~\bibnamefont{Hikasa}},
  \bibinfo{journal}{Nucl. Phys.} \textbf{\bibinfo{volume}{B282}},
  \bibinfo{pages}{253} (\bibinfo{year}{1987}).

\bibitem[{\citenamefont{Grzadkowski et~al.}(2010)\citenamefont{Grzadkowski,
  Iskrzynski, Misiak, and Rosiek}}]{Grzadkowski:2010es}
\bibinfo{author}{\bibfnamefont{B.}~\bibnamefont{Grzadkowski}},
  \bibinfo{author}{\bibfnamefont{M.}~\bibnamefont{Iskrzynski}},
  \bibinfo{author}{\bibfnamefont{M.}~\bibnamefont{Misiak}}, \bibnamefont{and}
  \bibinfo{author}{\bibfnamefont{J.}~\bibnamefont{Rosiek}},
  \bibinfo{journal}{JHEP} \textbf{\bibinfo{volume}{10}}, \bibinfo{pages}{085}
  (\bibinfo{year}{2010}), \eprint{1008.4884}.

\bibitem[{\citenamefont{Englert et~al.}(2020)\citenamefont{Englert, Galler, and
  White}}]{Englert:2019rga}
\bibinfo{author}{\bibfnamefont{C.}~\bibnamefont{Englert}},
  \bibinfo{author}{\bibfnamefont{P.}~\bibnamefont{Galler}}, \bibnamefont{and}
  \bibinfo{author}{\bibfnamefont{C.~D.} \bibnamefont{White}},
  \bibinfo{journal}{Phys. Rev.} \textbf{\bibinfo{volume}{D101}},
  \bibinfo{pages}{035035} (\bibinfo{year}{2020}), \eprint{1908.05588}.

\bibitem[{\citenamefont{Brown et~al.}(2020)\citenamefont{Brown, Englert,
  Galler, and Stylianou}}]{Brown:2020uwk}
\bibinfo{author}{\bibfnamefont{S.}~\bibnamefont{Brown}},
  \bibinfo{author}{\bibfnamefont{C.}~\bibnamefont{Englert}},
  \bibinfo{author}{\bibfnamefont{P.}~\bibnamefont{Galler}}, \bibnamefont{and}
  \bibinfo{author}{\bibfnamefont{P.}~\bibnamefont{Stylianou}},
  \bibinfo{journal}{Phys. Rev.} \textbf{\bibinfo{volume}{D102}},
  \bibinfo{pages}{075021} (\bibinfo{year}{2020}), \eprint{2006.09112}.

\bibitem[{\citenamefont{Buarque~Franzosi
  et~al.}(2017)\citenamefont{Buarque~Franzosi, Vryonidou, and
  Zhang}}]{BuarqueFranzosi:2017jrj}
\bibinfo{author}{\bibfnamefont{D.}~\bibnamefont{Buarque~Franzosi}},
  \bibinfo{author}{\bibfnamefont{E.}~\bibnamefont{Vryonidou}},
  \bibnamefont{and} \bibinfo{author}{\bibfnamefont{C.}~\bibnamefont{Zhang}},
  \bibinfo{journal}{JHEP} \textbf{\bibinfo{volume}{10}}, \bibinfo{pages}{096}
  (\bibinfo{year}{2017}), \eprint{1707.06760}.

\bibitem[{\citenamefont{Das~Bakshi
  et~al.}(2020{\natexlab{a}})\citenamefont{Das~Bakshi, Chakrabortty, Englert,
  Spannowsky, and Stylianou}}]{DasBakshi:2020ejz}
\bibinfo{author}{\bibfnamefont{S.}~\bibnamefont{Das~Bakshi}},
  \bibinfo{author}{\bibfnamefont{J.}~\bibnamefont{Chakrabortty}},
  \bibinfo{author}{\bibfnamefont{C.}~\bibnamefont{Englert}},
  \bibinfo{author}{\bibfnamefont{M.}~\bibnamefont{Spannowsky}},
  \bibnamefont{and} \bibinfo{author}{\bibfnamefont{P.}~\bibnamefont{Stylianou}}
  (\bibinfo{year}{2020}{\natexlab{a}}), \eprint{2009.13394}.

\bibitem[{\citenamefont{Aad et~al.}(2020{\natexlab{b}})}]{Aad:2020mkp}
\bibinfo{author}{\bibfnamefont{G.}~\bibnamefont{Aad}} \bibnamefont{et~al.}
  (\bibinfo{collaboration}{ATLAS}) (\bibinfo{year}{2020}{\natexlab{b}}),
  \eprint{2004.03447}.

\bibitem[{\citenamefont{Sirunyan
  et~al.}(2020{\natexlab{a}})}]{Sirunyan:2020tqm}
\bibinfo{author}{\bibfnamefont{A.~M.} \bibnamefont{Sirunyan}}
  \bibnamefont{et~al.} (\bibinfo{collaboration}{CMS})
  (\bibinfo{year}{2020}{\natexlab{a}}), \eprint{2012.04120}.

\bibitem[{\citenamefont{Sirunyan et~al.}(2021)}]{Sirunyan:2021zud}
\bibinfo{author}{\bibfnamefont{A.~M.} \bibnamefont{Sirunyan}}
  \bibnamefont{et~al.} (\bibinfo{collaboration}{CMS}) (\bibinfo{year}{2021}),
  \eprint{2102.02283}.

\bibitem[{\citenamefont{Plehn et~al.}(2002)\citenamefont{Plehn, Rainwater, and
  Zeppenfeld}}]{Plehn:2001nj}
\bibinfo{author}{\bibfnamefont{T.}~\bibnamefont{Plehn}},
  \bibinfo{author}{\bibfnamefont{D.~L.} \bibnamefont{Rainwater}},
  \bibnamefont{and}
  \bibinfo{author}{\bibfnamefont{D.}~\bibnamefont{Zeppenfeld}},
  \bibinfo{journal}{Phys. Rev. Lett.} \textbf{\bibinfo{volume}{88}},
  \bibinfo{pages}{051801} (\bibinfo{year}{2002}), \eprint{hep-ph/0105325}.

\bibitem[{\citenamefont{Hankele et~al.}(2006)\citenamefont{Hankele, Klamke,
  Zeppenfeld, and Figy}}]{Hankele:2006ma}
\bibinfo{author}{\bibfnamefont{V.}~\bibnamefont{Hankele}},
  \bibinfo{author}{\bibfnamefont{G.}~\bibnamefont{Klamke}},
  \bibinfo{author}{\bibfnamefont{D.}~\bibnamefont{Zeppenfeld}},
  \bibnamefont{and} \bibinfo{author}{\bibfnamefont{T.}~\bibnamefont{Figy}},
  \bibinfo{journal}{Phys. Rev.} \textbf{\bibinfo{volume}{D74}},
  \bibinfo{pages}{095001} (\bibinfo{year}{2006}), \eprint{hep-ph/0609075}.

\bibitem[{\citenamefont{Klamke and Zeppenfeld}(2007)}]{Klamke:2007cu}
\bibinfo{author}{\bibfnamefont{G.}~\bibnamefont{Klamke}} \bibnamefont{and}
  \bibinfo{author}{\bibfnamefont{D.}~\bibnamefont{Zeppenfeld}},
  \bibinfo{journal}{JHEP} \textbf{\bibinfo{volume}{04}}, \bibinfo{pages}{052}
  (\bibinfo{year}{2007}), \eprint{hep-ph/0703202}.

\bibitem[{\citenamefont{Campanario
  et~al.}(2011{\natexlab{a}})\citenamefont{Campanario, Kubocz, and
  Zeppenfeld}}]{Campanario:2010mi}
\bibinfo{author}{\bibfnamefont{F.}~\bibnamefont{Campanario}},
  \bibinfo{author}{\bibfnamefont{M.}~\bibnamefont{Kubocz}}, \bibnamefont{and}
  \bibinfo{author}{\bibfnamefont{D.}~\bibnamefont{Zeppenfeld}},
  \bibinfo{journal}{Phys. Rev.} \textbf{\bibinfo{volume}{D84}},
  \bibinfo{pages}{095025} (\bibinfo{year}{2011}{\natexlab{a}}),
  \eprint{1011.3819}.

\bibitem[{\citenamefont{Brehmer et~al.}(2018)\citenamefont{Brehmer, Kling,
  Plehn, and Tait}}]{Brehmer:2017lrt}
\bibinfo{author}{\bibfnamefont{J.}~\bibnamefont{Brehmer}},
  \bibinfo{author}{\bibfnamefont{F.}~\bibnamefont{Kling}},
  \bibinfo{author}{\bibfnamefont{T.}~\bibnamefont{Plehn}}, \bibnamefont{and}
  \bibinfo{author}{\bibfnamefont{T.~M.~P.} \bibnamefont{Tait}},
  \bibinfo{journal}{Phys. Rev.} \textbf{\bibinfo{volume}{D97}},
  \bibinfo{pages}{095017} (\bibinfo{year}{2018}), \eprint{1712.02350}.

\bibitem[{\citenamefont{Bernlochner et~al.}(2019)\citenamefont{Bernlochner,
  Englert, Hays, Lohwasser, Mildner, Pilkington, Price, and
  Spannowsky}}]{Bernlochner:2018opw}
\bibinfo{author}{\bibfnamefont{F.~U.} \bibnamefont{Bernlochner}},
  \bibinfo{author}{\bibfnamefont{C.}~\bibnamefont{Englert}},
  \bibinfo{author}{\bibfnamefont{C.}~\bibnamefont{Hays}},
  \bibinfo{author}{\bibfnamefont{K.}~\bibnamefont{Lohwasser}},
  \bibinfo{author}{\bibfnamefont{H.}~\bibnamefont{Mildner}},
  \bibinfo{author}{\bibfnamefont{A.}~\bibnamefont{Pilkington}},
  \bibinfo{author}{\bibfnamefont{D.~D.} \bibnamefont{Price}}, \bibnamefont{and}
  \bibinfo{author}{\bibfnamefont{M.}~\bibnamefont{Spannowsky}},
  \bibinfo{journal}{Phys. Lett.} \textbf{\bibinfo{volume}{B790}},
  \bibinfo{pages}{372} (\bibinfo{year}{2019}), \eprint{1808.06577}.

\bibitem[{\citenamefont{Englert et~al.}(2019)\citenamefont{Englert, Galler,
  Pilkington, and Spannowsky}}]{Englert:2019xhk}
\bibinfo{author}{\bibfnamefont{C.}~\bibnamefont{Englert}},
  \bibinfo{author}{\bibfnamefont{P.}~\bibnamefont{Galler}},
  \bibinfo{author}{\bibfnamefont{A.}~\bibnamefont{Pilkington}},
  \bibnamefont{and}
  \bibinfo{author}{\bibfnamefont{M.}~\bibnamefont{Spannowsky}},
  \bibinfo{journal}{Phys. Rev.} \textbf{\bibinfo{volume}{D99}},
  \bibinfo{pages}{095007} (\bibinfo{year}{2019}), \eprint{1901.05982}.

\bibitem[{\citenamefont{Cirigliano et~al.}(2019)\citenamefont{Cirigliano,
  Crivellin, Dekens, de~Vries, Hoferichter, and
  Mereghetti}}]{Cirigliano:2019vfc}
\bibinfo{author}{\bibfnamefont{V.}~\bibnamefont{Cirigliano}},
  \bibinfo{author}{\bibfnamefont{A.}~\bibnamefont{Crivellin}},
  \bibinfo{author}{\bibfnamefont{W.}~\bibnamefont{Dekens}},
  \bibinfo{author}{\bibfnamefont{J.}~\bibnamefont{de~Vries}},
  \bibinfo{author}{\bibfnamefont{M.}~\bibnamefont{Hoferichter}},
  \bibnamefont{and}
  \bibinfo{author}{\bibfnamefont{E.}~\bibnamefont{Mereghetti}},
  \bibinfo{journal}{Phys. Rev. Lett.} \textbf{\bibinfo{volume}{123}},
  \bibinfo{pages}{051801} (\bibinfo{year}{2019}), \eprint{1903.03625}.

\bibitem[{\citenamefont{Jegerlehner}(2001)}]{Jegerlehner:2000dz}
\bibinfo{author}{\bibfnamefont{F.}~\bibnamefont{Jegerlehner}},
  \bibinfo{journal}{Eur. Phys. J.} \textbf{\bibinfo{volume}{C18}},
  \bibinfo{pages}{673} (\bibinfo{year}{2001}), \eprint{hep-th/0005255}.

\bibitem[{\citenamefont{Chanowitz et~al.}(1979)\citenamefont{Chanowitz, Furman,
  and Hinchliffe}}]{Chanowitz:1979zu}
\bibinfo{author}{\bibfnamefont{M.~S.} \bibnamefont{Chanowitz}},
  \bibinfo{author}{\bibfnamefont{M.}~\bibnamefont{Furman}}, \bibnamefont{and}
  \bibinfo{author}{\bibfnamefont{I.}~\bibnamefont{Hinchliffe}},
  \bibinfo{journal}{Nucl. Phys.} \textbf{\bibinfo{volume}{B159}},
  \bibinfo{pages}{225} (\bibinfo{year}{1979}).

\bibitem[{\citenamefont{Ahmed et~al.}(2020)\citenamefont{Ahmed, Bernreuther,
  Chen, and Czakon}}]{Ahmed:2020kme}
\bibinfo{author}{\bibfnamefont{T.}~\bibnamefont{Ahmed}},
  \bibinfo{author}{\bibfnamefont{W.}~\bibnamefont{Bernreuther}},
  \bibinfo{author}{\bibfnamefont{L.}~\bibnamefont{Chen}}, \bibnamefont{and}
  \bibinfo{author}{\bibfnamefont{M.}~\bibnamefont{Czakon}},
  \bibinfo{journal}{JHEP} \textbf{\bibinfo{volume}{07}}, \bibinfo{pages}{159}
  (\bibinfo{year}{2020}), \eprint{2004.13753}.

\bibitem[{\citenamefont{Heller et~al.}(2020)\citenamefont{Heller, von
  Manteuffel, Schabinger, and Spiesberger}}]{Heller:2020owb}
\bibinfo{author}{\bibfnamefont{M.}~\bibnamefont{Heller}},
  \bibinfo{author}{\bibfnamefont{A.}~\bibnamefont{von Manteuffel}},
  \bibinfo{author}{\bibfnamefont{R.~M.} \bibnamefont{Schabinger}},
  \bibnamefont{and}
  \bibinfo{author}{\bibfnamefont{H.}~\bibnamefont{Spiesberger}}
  (\bibinfo{year}{2020}), \eprint{2012.05918}.

\bibitem[{\citenamefont{Angelescu and Huang}(2020)}]{Angelescu:2020yzf}
\bibinfo{author}{\bibfnamefont{A.}~\bibnamefont{Angelescu}} \bibnamefont{and}
  \bibinfo{author}{\bibfnamefont{P.}~\bibnamefont{Huang}}
  (\bibinfo{year}{2020}), \eprint{2006.16532}.

\bibitem[{\citenamefont{Miller et~al.}(2001)\citenamefont{Miller, Choi, Eberle,
  Muhlleitner, and Zerwas}}]{Miller:2001bi}
\bibinfo{author}{\bibfnamefont{D.~J.} \bibnamefont{Miller}},
  \bibinfo{author}{\bibfnamefont{S.~Y.} \bibnamefont{Choi}},
  \bibinfo{author}{\bibfnamefont{B.}~\bibnamefont{Eberle}},
  \bibinfo{author}{\bibfnamefont{M.~M.} \bibnamefont{Muhlleitner}},
  \bibnamefont{and} \bibinfo{author}{\bibfnamefont{P.~M.}
  \bibnamefont{Zerwas}}, pp. \bibinfo{pages}{1825--1834}
  (\bibinfo{year}{2001}), \eprint{hep-ph/0102023}.

\bibitem[{\citenamefont{Choi et~al.}(2003)\citenamefont{Choi, Miller,
  Muhlleitner, and Zerwas}}]{Choi:2002jk}
\bibinfo{author}{\bibfnamefont{S.~Y.} \bibnamefont{Choi}},
  \bibinfo{author}{\bibfnamefont{D.~J.} \bibnamefont{Miller}},
  \bibinfo{author}{\bibfnamefont{M.~M.} \bibnamefont{Muhlleitner}},
  \bibnamefont{and} \bibinfo{author}{\bibfnamefont{P.~M.}
  \bibnamefont{Zerwas}}, \bibinfo{journal}{Phys. Lett.}
  \textbf{\bibinfo{volume}{B553}}, \bibinfo{pages}{61} (\bibinfo{year}{2003}),
  \eprint{hep-ph/0210077}.

\bibitem[{\citenamefont{Accomando et~al.}(2006)}]{Accomando:2006ga}
\bibinfo{author}{\bibfnamefont{E.}~\bibnamefont{Accomando}}
  \bibnamefont{et~al.} (\bibinfo{year}{2006}), \eprint{hep-ph/0608079}.

\bibitem[{\citenamefont{Barger et~al.}(2000)\citenamefont{Barger, Han, Li, and
  Plehn}}]{Barger:1999tn}
\bibinfo{author}{\bibfnamefont{V.~D.} \bibnamefont{Barger}},
  \bibinfo{author}{\bibfnamefont{T.}~\bibnamefont{Han}},
  \bibinfo{author}{\bibfnamefont{T.-J.} \bibnamefont{Li}}, \bibnamefont{and}
  \bibinfo{author}{\bibfnamefont{T.}~\bibnamefont{Plehn}},
  \bibinfo{journal}{Phys. Lett.} \textbf{\bibinfo{volume}{B475}},
  \bibinfo{pages}{342} (\bibinfo{year}{2000}), \eprint{hep-ph/9907425}.

\bibitem[{\citenamefont{Englert et~al.}(2013)\citenamefont{Englert,
  Goncalves-Netto, Mawatari, and Plehn}}]{Englert:2012xt}
\bibinfo{author}{\bibfnamefont{C.}~\bibnamefont{Englert}},
  \bibinfo{author}{\bibfnamefont{D.}~\bibnamefont{Goncalves-Netto}},
  \bibinfo{author}{\bibfnamefont{K.}~\bibnamefont{Mawatari}}, \bibnamefont{and}
  \bibinfo{author}{\bibfnamefont{T.}~\bibnamefont{Plehn}},
  \bibinfo{journal}{JHEP} \textbf{\bibinfo{volume}{01}}, \bibinfo{pages}{148}
  (\bibinfo{year}{2013}), \eprint{1212.0843}.

\bibitem[{\citenamefont{Englert et~al.}(2012)\citenamefont{Englert, Spannowsky,
  and Takeuchi}}]{Englert:2012ct}
\bibinfo{author}{\bibfnamefont{C.}~\bibnamefont{Englert}},
  \bibinfo{author}{\bibfnamefont{M.}~\bibnamefont{Spannowsky}},
  \bibnamefont{and} \bibinfo{author}{\bibfnamefont{M.}~\bibnamefont{Takeuchi}},
  \bibinfo{journal}{JHEP} \textbf{\bibinfo{volume}{06}}, \bibinfo{pages}{108}
  (\bibinfo{year}{2012}), \eprint{1203.5788}.

\bibitem[{\citenamefont{Dolan et~al.}(2014)\citenamefont{Dolan, Harris,
  Jankowiak, and Spannowsky}}]{Dolan:2014upa}
\bibinfo{author}{\bibfnamefont{M.~J.} \bibnamefont{Dolan}},
  \bibinfo{author}{\bibfnamefont{P.}~\bibnamefont{Harris}},
  \bibinfo{author}{\bibfnamefont{M.}~\bibnamefont{Jankowiak}},
  \bibnamefont{and}
  \bibinfo{author}{\bibfnamefont{M.}~\bibnamefont{Spannowsky}},
  \bibinfo{journal}{Phys. Rev.} \textbf{\bibinfo{volume}{D90}},
  \bibinfo{pages}{073008} (\bibinfo{year}{2014}), \eprint{1406.3322}.

\bibitem[{\citenamefont{Araz et~al.}(2020)\citenamefont{Araz, Banerjee, Gupta,
  and Spannowsky}}]{Araz:2020zyh}
\bibinfo{author}{\bibfnamefont{J.~Y.} \bibnamefont{Araz}},
  \bibinfo{author}{\bibfnamefont{S.}~\bibnamefont{Banerjee}},
  \bibinfo{author}{\bibfnamefont{R.~S.} \bibnamefont{Gupta}}, \bibnamefont{and}
  \bibinfo{author}{\bibfnamefont{M.}~\bibnamefont{Spannowsky}}
  (\bibinfo{year}{2020}), \eprint{2011.03555}.

\bibitem[{\citenamefont{Ethier et~al.}(2021)\citenamefont{Ethier,
  Gomez-Ambrosio, Magni, and Rojo}}]{Ethier:2021ydt}
\bibinfo{author}{\bibfnamefont{J.~J.} \bibnamefont{Ethier}},
  \bibinfo{author}{\bibfnamefont{R.}~\bibnamefont{Gomez-Ambrosio}},
  \bibinfo{author}{\bibfnamefont{G.}~\bibnamefont{Magni}}, \bibnamefont{and}
  \bibinfo{author}{\bibfnamefont{J.}~\bibnamefont{Rojo}}
  (\bibinfo{year}{2021}), \eprint{2101.03180}.

\bibitem[{\citenamefont{Franceschini et~al.}(2018)\citenamefont{Franceschini,
  Panico, Pomarol, Riva, and Wulzer}}]{Franceschini:2017xkh}
\bibinfo{author}{\bibfnamefont{R.}~\bibnamefont{Franceschini}},
  \bibinfo{author}{\bibfnamefont{G.}~\bibnamefont{Panico}},
  \bibinfo{author}{\bibfnamefont{A.}~\bibnamefont{Pomarol}},
  \bibinfo{author}{\bibfnamefont{F.}~\bibnamefont{Riva}}, \bibnamefont{and}
  \bibinfo{author}{\bibfnamefont{A.}~\bibnamefont{Wulzer}},
  \bibinfo{journal}{JHEP} \textbf{\bibinfo{volume}{02}}, \bibinfo{pages}{111}
  (\bibinfo{year}{2018}), \eprint{1712.01310}.

\bibitem[{\citenamefont{Banerjee
  et~al.}(2020{\natexlab{a}})\citenamefont{Banerjee, Gupta, Reiness, Seth, and
  Spannowsky}}]{Banerjee:2019twi}
\bibinfo{author}{\bibfnamefont{S.}~\bibnamefont{Banerjee}},
  \bibinfo{author}{\bibfnamefont{R.~S.} \bibnamefont{Gupta}},
  \bibinfo{author}{\bibfnamefont{J.~Y.} \bibnamefont{Reiness}},
  \bibinfo{author}{\bibfnamefont{S.}~\bibnamefont{Seth}}, \bibnamefont{and}
  \bibinfo{author}{\bibfnamefont{M.}~\bibnamefont{Spannowsky}},
  \bibinfo{journal}{JHEP} \textbf{\bibinfo{volume}{09}}, \bibinfo{pages}{170}
  (\bibinfo{year}{2020}{\natexlab{a}}), \eprint{1912.07628}.

\bibitem[{\citenamefont{Banerjee
  et~al.}(2020{\natexlab{b}})\citenamefont{Banerjee, Gupta, Ochoa-Valeriano,
  Spannowsky, and Venturini}}]{Banerjee:2020vtm}
\bibinfo{author}{\bibfnamefont{S.}~\bibnamefont{Banerjee}},
  \bibinfo{author}{\bibfnamefont{R.~S.} \bibnamefont{Gupta}},
  \bibinfo{author}{\bibfnamefont{O.}~\bibnamefont{Ochoa-Valeriano}},
  \bibinfo{author}{\bibfnamefont{M.}~\bibnamefont{Spannowsky}},
  \bibnamefont{and} \bibinfo{author}{\bibfnamefont{E.}~\bibnamefont{Venturini}}
  (\bibinfo{year}{2020}{\natexlab{b}}), \eprint{2012.11631}.

\bibitem[{\citenamefont{Khachatryan et~al.}(2015)}]{Khachatryan:2015bnx}
\bibinfo{author}{\bibfnamefont{V.}~\bibnamefont{Khachatryan}}
  \bibnamefont{et~al.} (\bibinfo{collaboration}{CMS}), \bibinfo{journal}{Phys.
  Rev.} \textbf{\bibinfo{volume}{D92}}, \bibinfo{pages}{032008}
  (\bibinfo{year}{2015}), \eprint{1506.01010}.

\bibitem[{\citenamefont{Englert et~al.}(2016)\citenamefont{Englert, Mattelaer,
  and Spannowsky}}]{Englert:2015dlp}
\bibinfo{author}{\bibfnamefont{C.}~\bibnamefont{Englert}},
  \bibinfo{author}{\bibfnamefont{O.}~\bibnamefont{Mattelaer}},
  \bibnamefont{and}
  \bibinfo{author}{\bibfnamefont{M.}~\bibnamefont{Spannowsky}},
  \bibinfo{journal}{Phys. Lett.} \textbf{\bibinfo{volume}{B756}},
  \bibinfo{pages}{103} (\bibinfo{year}{2016}), \eprint{1512.03429}.

\bibitem[{\citenamefont{Ngairangbam et~al.}(2020)\citenamefont{Ngairangbam,
  Bhardwaj, Konar, and Nayak}}]{Ngairangbam:2020ksz}
\bibinfo{author}{\bibfnamefont{V.~S.} \bibnamefont{Ngairangbam}},
  \bibinfo{author}{\bibfnamefont{A.}~\bibnamefont{Bhardwaj}},
  \bibinfo{author}{\bibfnamefont{P.}~\bibnamefont{Konar}}, \bibnamefont{and}
  \bibinfo{author}{\bibfnamefont{A.~K.} \bibnamefont{Nayak}},
  \bibinfo{journal}{Eur. Phys. J.} \textbf{\bibinfo{volume}{C80}},
  \bibinfo{pages}{1055} (\bibinfo{year}{2020}), \eprint{2008.05434}.

\bibitem[{\citenamefont{Andersen et~al.}(2013)\citenamefont{Andersen, Englert,
  and Spannowsky}}]{Andersen:2012kn}
\bibinfo{author}{\bibfnamefont{J.~R.} \bibnamefont{Andersen}},
  \bibinfo{author}{\bibfnamefont{C.}~\bibnamefont{Englert}}, \bibnamefont{and}
  \bibinfo{author}{\bibfnamefont{M.}~\bibnamefont{Spannowsky}},
  \bibinfo{journal}{Phys. Rev.} \textbf{\bibinfo{volume}{D87}},
  \bibinfo{pages}{015019} (\bibinfo{year}{2013}), \eprint{1211.3011}.

\bibitem[{\citenamefont{Christensen and Duhr}(2009)}]{Christensen:2008py}
\bibinfo{author}{\bibfnamefont{N.~D.} \bibnamefont{Christensen}}
  \bibnamefont{and} \bibinfo{author}{\bibfnamefont{C.}~\bibnamefont{Duhr}},
  \bibinfo{journal}{Comput. Phys. Commun.} \textbf{\bibinfo{volume}{180}},
  \bibinfo{pages}{1614} (\bibinfo{year}{2009}), \eprint{0806.4194}.

\bibitem[{\citenamefont{Alloul et~al.}(2014)\citenamefont{Alloul, Christensen,
  Degrande, Duhr, and Fuks}}]{Alloul:2013bka}
\bibinfo{author}{\bibfnamefont{A.}~\bibnamefont{Alloul}},
  \bibinfo{author}{\bibfnamefont{N.~D.} \bibnamefont{Christensen}},
  \bibinfo{author}{\bibfnamefont{C.}~\bibnamefont{Degrande}},
  \bibinfo{author}{\bibfnamefont{C.}~\bibnamefont{Duhr}}, \bibnamefont{and}
  \bibinfo{author}{\bibfnamefont{B.}~\bibnamefont{Fuks}},
  \bibinfo{journal}{Comput. Phys. Commun.} \textbf{\bibinfo{volume}{185}},
  \bibinfo{pages}{2250} (\bibinfo{year}{2014}), \eprint{1310.1921}.

\bibitem[{\citenamefont{Degrande et~al.}(2012)\citenamefont{Degrande, Duhr,
  Fuks, Grellscheid, Mattelaer, and Reiter}}]{Degrande:2011ua}
\bibinfo{author}{\bibfnamefont{C.}~\bibnamefont{Degrande}},
  \bibinfo{author}{\bibfnamefont{C.}~\bibnamefont{Duhr}},
  \bibinfo{author}{\bibfnamefont{B.}~\bibnamefont{Fuks}},
  \bibinfo{author}{\bibfnamefont{D.}~\bibnamefont{Grellscheid}},
  \bibinfo{author}{\bibfnamefont{O.}~\bibnamefont{Mattelaer}},
  \bibnamefont{and} \bibinfo{author}{\bibfnamefont{T.}~\bibnamefont{Reiter}},
  \bibinfo{journal}{Comput. Phys. Commun.} \textbf{\bibinfo{volume}{183}},
  \bibinfo{pages}{1201} (\bibinfo{year}{2012}), \eprint{1108.2040}.

\bibitem[{\citenamefont{Alwall et~al.}(2014)\citenamefont{Alwall, Frederix,
  Frixione, Hirschi, Maltoni, Mattelaer, Shao, Stelzer, Torrielli, and
  Zaro}}]{Alwall:2014hca}
\bibinfo{author}{\bibfnamefont{J.}~\bibnamefont{Alwall}},
  \bibinfo{author}{\bibfnamefont{R.}~\bibnamefont{Frederix}},
  \bibinfo{author}{\bibfnamefont{S.}~\bibnamefont{Frixione}},
  \bibinfo{author}{\bibfnamefont{V.}~\bibnamefont{Hirschi}},
  \bibinfo{author}{\bibfnamefont{F.}~\bibnamefont{Maltoni}},
  \bibinfo{author}{\bibfnamefont{O.}~\bibnamefont{Mattelaer}},
  \bibinfo{author}{\bibfnamefont{H.~S.} \bibnamefont{Shao}},
  \bibinfo{author}{\bibfnamefont{T.}~\bibnamefont{Stelzer}},
  \bibinfo{author}{\bibfnamefont{P.}~\bibnamefont{Torrielli}},
  \bibnamefont{and} \bibinfo{author}{\bibfnamefont{M.}~\bibnamefont{Zaro}},
  \bibinfo{journal}{JHEP} \textbf{\bibinfo{volume}{07}}, \bibinfo{pages}{079}
  (\bibinfo{year}{2014}), \eprint{1405.0301}.

\bibitem[{\citenamefont{Alwall et~al.}(2011)\citenamefont{Alwall, Herquet,
  Maltoni, Mattelaer, and Stelzer}}]{Alwall:2011uj}
\bibinfo{author}{\bibfnamefont{J.}~\bibnamefont{Alwall}},
  \bibinfo{author}{\bibfnamefont{M.}~\bibnamefont{Herquet}},
  \bibinfo{author}{\bibfnamefont{F.}~\bibnamefont{Maltoni}},
  \bibinfo{author}{\bibfnamefont{O.}~\bibnamefont{Mattelaer}},
  \bibnamefont{and} \bibinfo{author}{\bibfnamefont{T.}~\bibnamefont{Stelzer}},
  \bibinfo{journal}{JHEP} \textbf{\bibinfo{volume}{06}}, \bibinfo{pages}{128}
  (\bibinfo{year}{2011}), \eprint{1106.0522}.

\bibitem[{\citenamefont{de~Aquino et~al.}(2012)\citenamefont{de~Aquino, Link,
  Maltoni, Mattelaer, and Stelzer}}]{deAquino:2011ub}
\bibinfo{author}{\bibfnamefont{P.}~\bibnamefont{de~Aquino}},
  \bibinfo{author}{\bibfnamefont{W.}~\bibnamefont{Link}},
  \bibinfo{author}{\bibfnamefont{F.}~\bibnamefont{Maltoni}},
  \bibinfo{author}{\bibfnamefont{O.}~\bibnamefont{Mattelaer}},
  \bibnamefont{and} \bibinfo{author}{\bibfnamefont{T.}~\bibnamefont{Stelzer}},
  \bibinfo{journal}{Comput. Phys. Commun.} \textbf{\bibinfo{volume}{183}},
  \bibinfo{pages}{2254} (\bibinfo{year}{2012}), \eprint{1108.2041}.

\bibitem[{\citenamefont{Dokshitzer et~al.}(1987)\citenamefont{Dokshitzer,
  Troian, and Khoze}}]{Dokshitzer:1987nc}
\bibinfo{author}{\bibfnamefont{Y.~L.} \bibnamefont{Dokshitzer}},
  \bibinfo{author}{\bibfnamefont{S.~I.} \bibnamefont{Troian}},
  \bibnamefont{and} \bibinfo{author}{\bibfnamefont{V.~A.} \bibnamefont{Khoze}},
  \bibinfo{journal}{Sov. J. Nucl. Phys.} \textbf{\bibinfo{volume}{46}},
  \bibinfo{pages}{712} (\bibinfo{year}{1987}), \bibinfo{note}{[Yad.
  Fiz.46,1220(1987)]}.

\bibitem[{\citenamefont{Dokshitzer et~al.}(1992)\citenamefont{Dokshitzer,
  Khoze, and Sjostrand}}]{Dokshitzer:1991he}
\bibinfo{author}{\bibfnamefont{Y.~L.} \bibnamefont{Dokshitzer}},
  \bibinfo{author}{\bibfnamefont{V.~A.} \bibnamefont{Khoze}}, \bibnamefont{and}
  \bibinfo{author}{\bibfnamefont{T.}~\bibnamefont{Sjostrand}},
  \bibinfo{journal}{Phys. Lett.} \textbf{\bibinfo{volume}{B274}},
  \bibinfo{pages}{116} (\bibinfo{year}{1992}).

\bibitem[{\citenamefont{Barger et~al.}(1991)\citenamefont{Barger, Cheung, Han,
  and Zeppenfeld}}]{Barger:1991ar}
\bibinfo{author}{\bibfnamefont{V.~D.} \bibnamefont{Barger}},
  \bibinfo{author}{\bibfnamefont{K.-m.} \bibnamefont{Cheung}},
  \bibinfo{author}{\bibfnamefont{T.}~\bibnamefont{Han}}, \bibnamefont{and}
  \bibinfo{author}{\bibfnamefont{D.}~\bibnamefont{Zeppenfeld}},
  \bibinfo{journal}{Phys. Rev.} \textbf{\bibinfo{volume}{D44}},
  \bibinfo{pages}{2701} (\bibinfo{year}{1991}), \bibinfo{note}{[Erratum: Phys.
  Rev.D48,5444(1993)]}.

\bibitem[{\citenamefont{Bjorken}(1993)}]{Bjorken:1992er}
\bibinfo{author}{\bibfnamefont{J.~D.} \bibnamefont{Bjorken}},
  \bibinfo{journal}{Phys. Rev.} \textbf{\bibinfo{volume}{D47}},
  \bibinfo{pages}{101} (\bibinfo{year}{1993}).

\bibitem[{\citenamefont{Khachatryan et~al.}(2020)}]{CMS:2020olm}
\bibinfo{author}{\bibfnamefont{V.}~\bibnamefont{Khachatryan}}
  \bibnamefont{et~al.} (\bibinfo{collaboration}{CMS}) (\bibinfo{year}{2020}),
  \eprint{CMS-PAS-SMP-19-002}.

\bibitem[{\citenamefont{Sjöstrand et~al.}(2015)\citenamefont{Sjöstrand, Ask,
  Christiansen, Corke, Desai, Ilten, Mrenna, Prestel, Rasmussen, and
  Skands}}]{Sjostrand:2014zea}
\bibinfo{author}{\bibfnamefont{T.}~\bibnamefont{Sjöstrand}},
  \bibinfo{author}{\bibfnamefont{S.}~\bibnamefont{Ask}},
  \bibinfo{author}{\bibfnamefont{J.~R.} \bibnamefont{Christiansen}},
  \bibinfo{author}{\bibfnamefont{R.}~\bibnamefont{Corke}},
  \bibinfo{author}{\bibfnamefont{N.}~\bibnamefont{Desai}},
  \bibinfo{author}{\bibfnamefont{P.}~\bibnamefont{Ilten}},
  \bibinfo{author}{\bibfnamefont{S.}~\bibnamefont{Mrenna}},
  \bibinfo{author}{\bibfnamefont{S.}~\bibnamefont{Prestel}},
  \bibinfo{author}{\bibfnamefont{C.~O.} \bibnamefont{Rasmussen}},
  \bibnamefont{and} \bibinfo{author}{\bibfnamefont{P.~Z.}
  \bibnamefont{Skands}}, \bibinfo{journal}{Comput. Phys. Commun.}
  \textbf{\bibinfo{volume}{191}}, \bibinfo{pages}{159} (\bibinfo{year}{2015}),
  \eprint{1410.3012}.

\bibitem[{\citenamefont{Figy et~al.}(2003)\citenamefont{Figy, Oleari, and
  Zeppenfeld}}]{Figy:2003nv}
\bibinfo{author}{\bibfnamefont{T.}~\bibnamefont{Figy}},
  \bibinfo{author}{\bibfnamefont{C.}~\bibnamefont{Oleari}}, \bibnamefont{and}
  \bibinfo{author}{\bibfnamefont{D.}~\bibnamefont{Zeppenfeld}},
  \bibinfo{journal}{Phys. Rev.} \textbf{\bibinfo{volume}{D68}},
  \bibinfo{pages}{073005} (\bibinfo{year}{2003}), \eprint{hep-ph/0306109}.

\bibitem[{\citenamefont{Conte et~al.}(2013)\citenamefont{Conte, Fuks, and
  Serret}}]{Conte:2012fm}
\bibinfo{author}{\bibfnamefont{E.}~\bibnamefont{Conte}},
  \bibinfo{author}{\bibfnamefont{B.}~\bibnamefont{Fuks}}, \bibnamefont{and}
  \bibinfo{author}{\bibfnamefont{G.}~\bibnamefont{Serret}},
  \bibinfo{journal}{Comput. Phys. Commun.} \textbf{\bibinfo{volume}{184}},
  \bibinfo{pages}{222} (\bibinfo{year}{2013}), \eprint{1206.1599}.

\bibitem[{\citenamefont{Conte et~al.}(2014)\citenamefont{Conte, Dumont, Fuks,
  and Wymant}}]{Conte:2014zja}
\bibinfo{author}{\bibfnamefont{E.}~\bibnamefont{Conte}},
  \bibinfo{author}{\bibfnamefont{B.}~\bibnamefont{Dumont}},
  \bibinfo{author}{\bibfnamefont{B.}~\bibnamefont{Fuks}}, \bibnamefont{and}
  \bibinfo{author}{\bibfnamefont{C.}~\bibnamefont{Wymant}},
  \bibinfo{journal}{Eur. Phys. J.} \textbf{\bibinfo{volume}{C74}},
  \bibinfo{pages}{3103} (\bibinfo{year}{2014}), \eprint{1405.3982}.

\bibitem[{\citenamefont{Dumont et~al.}(2015)\citenamefont{Dumont, Fuks, Kraml,
  Bein, Chalons, Conte, Kulkarni, Sengupta, and Wymant}}]{Dumont:2014tja}
\bibinfo{author}{\bibfnamefont{B.}~\bibnamefont{Dumont}},
  \bibinfo{author}{\bibfnamefont{B.}~\bibnamefont{Fuks}},
  \bibinfo{author}{\bibfnamefont{S.}~\bibnamefont{Kraml}},
  \bibinfo{author}{\bibfnamefont{S.}~\bibnamefont{Bein}},
  \bibinfo{author}{\bibfnamefont{G.}~\bibnamefont{Chalons}},
  \bibinfo{author}{\bibfnamefont{E.}~\bibnamefont{Conte}},
  \bibinfo{author}{\bibfnamefont{S.}~\bibnamefont{Kulkarni}},
  \bibinfo{author}{\bibfnamefont{D.}~\bibnamefont{Sengupta}}, \bibnamefont{and}
  \bibinfo{author}{\bibfnamefont{C.}~\bibnamefont{Wymant}},
  \bibinfo{journal}{Eur. Phys. J.} \textbf{\bibinfo{volume}{C75}},
  \bibinfo{pages}{56} (\bibinfo{year}{2015}), \eprint{1407.3278}.

\bibitem[{\citenamefont{Conte and Fuks}(2018)}]{Conte:2018vmg}
\bibinfo{author}{\bibfnamefont{E.}~\bibnamefont{Conte}} \bibnamefont{and}
  \bibinfo{author}{\bibfnamefont{B.}~\bibnamefont{Fuks}},
  \bibinfo{journal}{Int. J. Mod. Phys.} \textbf{\bibinfo{volume}{A33}},
  \bibinfo{pages}{1830027} (\bibinfo{year}{2018}), \eprint{1808.00480}.

\bibitem[{\citenamefont{Cacciari et~al.}(2012)\citenamefont{Cacciari, Salam,
  and Soyez}}]{Cacciari:2011ma}
\bibinfo{author}{\bibfnamefont{M.}~\bibnamefont{Cacciari}},
  \bibinfo{author}{\bibfnamefont{G.~P.} \bibnamefont{Salam}}, \bibnamefont{and}
  \bibinfo{author}{\bibfnamefont{G.}~\bibnamefont{Soyez}},
  \bibinfo{journal}{Eur. Phys. J.} \textbf{\bibinfo{volume}{C72}},
  \bibinfo{pages}{1896} (\bibinfo{year}{2012}), \eprint{1111.6097}.

\bibitem[{\citenamefont{Cacciari and Salam}(2006)}]{Cacciari:2005hq}
\bibinfo{author}{\bibfnamefont{M.}~\bibnamefont{Cacciari}} \bibnamefont{and}
  \bibinfo{author}{\bibfnamefont{G.~P.} \bibnamefont{Salam}},
  \bibinfo{journal}{Phys. Lett.} \textbf{\bibinfo{volume}{B641}},
  \bibinfo{pages}{57} (\bibinfo{year}{2006}), \eprint{hep-ph/0512210}.

\bibitem[{\citenamefont{Cacciari et~al.}(2008)\citenamefont{Cacciari, Salam,
  and Soyez}}]{Cacciari:2008gp}
\bibinfo{author}{\bibfnamefont{M.}~\bibnamefont{Cacciari}},
  \bibinfo{author}{\bibfnamefont{G.~P.} \bibnamefont{Salam}}, \bibnamefont{and}
  \bibinfo{author}{\bibfnamefont{G.}~\bibnamefont{Soyez}},
  \bibinfo{journal}{JHEP} \textbf{\bibinfo{volume}{04}}, \bibinfo{pages}{063}
  (\bibinfo{year}{2008}), \eprint{0802.1189}.

\bibitem[{\citenamefont{Samuel}(1983)}]{Samuel:1983eg}
\bibinfo{author}{\bibfnamefont{M.~A.} \bibnamefont{Samuel}},
  \bibinfo{journal}{Phys. Rev.} \textbf{\bibinfo{volume}{D27}},
  \bibinfo{pages}{2724} (\bibinfo{year}{1983}).

\bibitem[{\citenamefont{Samuel et~al.}(1984)\citenamefont{Samuel, Sen,
  Sylvester, and Laursen}}]{Samuel:1984ru}
\bibinfo{author}{\bibfnamefont{M.~A.} \bibnamefont{Samuel}},
  \bibinfo{author}{\bibfnamefont{A.}~\bibnamefont{Sen}},
  \bibinfo{author}{\bibfnamefont{G.~S.} \bibnamefont{Sylvester}},
  \bibnamefont{and} \bibinfo{author}{\bibfnamefont{M.~L.}
  \bibnamefont{Laursen}}, \bibinfo{journal}{Phys. Rev.}
  \textbf{\bibinfo{volume}{D29}}, \bibinfo{pages}{994} (\bibinfo{year}{1984}).

\bibitem[{\citenamefont{Brodsky and Brown}(1982)}]{Brodsky:1982sh}
\bibinfo{author}{\bibfnamefont{S.~J.} \bibnamefont{Brodsky}} \bibnamefont{and}
  \bibinfo{author}{\bibfnamefont{R.~W.} \bibnamefont{Brown}},
  \bibinfo{journal}{Phys. Rev. Lett.} \textbf{\bibinfo{volume}{49}},
  \bibinfo{pages}{966} (\bibinfo{year}{1982}).

\bibitem[{\citenamefont{Brown et~al.}(1983)\citenamefont{Brown, Kowalski, and
  Brodsky}}]{Brown:1982xx}
\bibinfo{author}{\bibfnamefont{R.~W.} \bibnamefont{Brown}},
  \bibinfo{author}{\bibfnamefont{K.~L.} \bibnamefont{Kowalski}},
  \bibnamefont{and} \bibinfo{author}{\bibfnamefont{S.~J.}
  \bibnamefont{Brodsky}}, \bibinfo{journal}{Phys. Rev.}
  \textbf{\bibinfo{volume}{D28}}, \bibinfo{pages}{624} (\bibinfo{year}{1983}),
  \bibinfo{note}{[Addendum: Phys. Rev.D29,2100(1984)]}.

\bibitem[{\citenamefont{Diakonos et~al.}(1993)\citenamefont{Diakonos,
  Korakianitis, Papadopoulos, Philippides, and Stirling}}]{Diakonos:1992qc}
\bibinfo{author}{\bibfnamefont{F.~K.} \bibnamefont{Diakonos}},
  \bibinfo{author}{\bibfnamefont{O.}~\bibnamefont{Korakianitis}},
  \bibinfo{author}{\bibfnamefont{C.~G.} \bibnamefont{Papadopoulos}},
  \bibinfo{author}{\bibfnamefont{C.}~\bibnamefont{Philippides}},
  \bibnamefont{and} \bibinfo{author}{\bibfnamefont{W.~J.}
  \bibnamefont{Stirling}}, \bibinfo{journal}{Phys. Lett.}
  \textbf{\bibinfo{volume}{B303}}, \bibinfo{pages}{177} (\bibinfo{year}{1993}),
  \eprint{hep-ph/9301238}.

\bibitem[{\citenamefont{Baur et~al.}(1993)\citenamefont{Baur, Han, and
  Ohnemus}}]{Baur:1993ir}
\bibinfo{author}{\bibfnamefont{U.}~\bibnamefont{Baur}},
  \bibinfo{author}{\bibfnamefont{T.}~\bibnamefont{Han}}, \bibnamefont{and}
  \bibinfo{author}{\bibfnamefont{J.}~\bibnamefont{Ohnemus}},
  \bibinfo{journal}{Phys. Rev.} \textbf{\bibinfo{volume}{D48}},
  \bibinfo{pages}{5140} (\bibinfo{year}{1993}), \eprint{hep-ph/9305314}.

\bibitem[{\citenamefont{Campanario
  et~al.}(2011{\natexlab{b}})\citenamefont{Campanario, Englert, and
  Spannowsky}}]{Campanario:2010hv}
\bibinfo{author}{\bibfnamefont{F.}~\bibnamefont{Campanario}},
  \bibinfo{author}{\bibfnamefont{C.}~\bibnamefont{Englert}}, \bibnamefont{and}
  \bibinfo{author}{\bibfnamefont{M.}~\bibnamefont{Spannowsky}},
  \bibinfo{journal}{Phys. Rev.} \textbf{\bibinfo{volume}{D83}},
  \bibinfo{pages}{074009} (\bibinfo{year}{2011}{\natexlab{b}}),
  \eprint{1010.1291}.

\bibitem[{\citenamefont{Aaboud et~al.}(2019)}]{Aaboud:2019nkz}
\bibinfo{author}{\bibfnamefont{M.}~\bibnamefont{Aaboud}} \bibnamefont{et~al.}
  (\bibinfo{collaboration}{ATLAS}), \bibinfo{journal}{Eur. Phys. J.}
  \textbf{\bibinfo{volume}{C79}}, \bibinfo{pages}{884} (\bibinfo{year}{2019}),
  \eprint{1905.04242}.

\bibitem[{\citenamefont{Khachatryan et~al.}(2017)}]{Khachatryan:2016tgp}
\bibinfo{author}{\bibfnamefont{V.}~\bibnamefont{Khachatryan}}
  \bibnamefont{et~al.} (\bibinfo{collaboration}{CMS}), \bibinfo{journal}{Phys.
  Lett.} \textbf{\bibinfo{volume}{B766}}, \bibinfo{pages}{268}
  (\bibinfo{year}{2017}), \eprint{1607.06943}.

\bibitem[{\citenamefont{Peskin and Takeuchi}(1992)}]{Peskin:1991sw}
\bibinfo{author}{\bibfnamefont{M.~E.} \bibnamefont{Peskin}} \bibnamefont{and}
  \bibinfo{author}{\bibfnamefont{T.}~\bibnamefont{Takeuchi}},
  \bibinfo{journal}{Phys. Rev.} \textbf{\bibinfo{volume}{D46}},
  \bibinfo{pages}{381} (\bibinfo{year}{1992}).

\bibitem[{\citenamefont{Peskin and Takeuchi}(1990)}]{Peskin:1990zt}
\bibinfo{author}{\bibfnamefont{M.~E.} \bibnamefont{Peskin}} \bibnamefont{and}
  \bibinfo{author}{\bibfnamefont{T.}~\bibnamefont{Takeuchi}},
  \bibinfo{journal}{Phys. Rev. Lett.} \textbf{\bibinfo{volume}{65}},
  \bibinfo{pages}{964} (\bibinfo{year}{1990}).

\bibitem[{\citenamefont{Baak et~al.}(2014)\citenamefont{Baak, Cúth, Haller,
  Hoecker, Kogler, Mönig, Schott, and Stelzer}}]{Baak:2014ora}
\bibinfo{author}{\bibfnamefont{M.}~\bibnamefont{Baak}},
  \bibinfo{author}{\bibfnamefont{J.}~\bibnamefont{Cúth}},
  \bibinfo{author}{\bibfnamefont{J.}~\bibnamefont{Haller}},
  \bibinfo{author}{\bibfnamefont{A.}~\bibnamefont{Hoecker}},
  \bibinfo{author}{\bibfnamefont{R.}~\bibnamefont{Kogler}},
  \bibinfo{author}{\bibfnamefont{K.}~\bibnamefont{Mönig}},
  \bibinfo{author}{\bibfnamefont{M.}~\bibnamefont{Schott}}, \bibnamefont{and}
  \bibinfo{author}{\bibfnamefont{J.}~\bibnamefont{Stelzer}}
  (\bibinfo{collaboration}{Gfitter Group}), \bibinfo{journal}{Eur. Phys. J.}
  \textbf{\bibinfo{volume}{C74}}, \bibinfo{pages}{3046} (\bibinfo{year}{2014}),
  \eprint{1407.3792}.

\bibitem[{\citenamefont{Kalinowski et~al.}(2018)\citenamefont{Kalinowski,
  Koz\'ow, Pokorski, Rosiek, Szleper, and Tkaczyk}}]{Kalinowski:2018oxd}
\bibinfo{author}{\bibfnamefont{J.}~\bibnamefont{Kalinowski}},
  \bibinfo{author}{\bibfnamefont{P.}~\bibnamefont{Koz\'ow}},
  \bibinfo{author}{\bibfnamefont{S.}~\bibnamefont{Pokorski}},
  \bibinfo{author}{\bibfnamefont{J.}~\bibnamefont{Rosiek}},
  \bibinfo{author}{\bibfnamefont{M.}~\bibnamefont{Szleper}}, \bibnamefont{and}
  \bibinfo{author}{\bibfnamefont{S.}~\bibnamefont{Tkaczyk}},
  \bibinfo{journal}{Eur. Phys. J. C} \textbf{\bibinfo{volume}{78}},
  \bibinfo{pages}{403} (\bibinfo{year}{2018}), \eprint{1802.02366}.

\bibitem[{\citenamefont{Koz\'ow et~al.}(2019)\citenamefont{Koz\'ow, Merlo,
  Pokorski, and Szleper}}]{Kozow:2019txg}
\bibinfo{author}{\bibfnamefont{P.}~\bibnamefont{Koz\'ow}},
  \bibinfo{author}{\bibfnamefont{L.}~\bibnamefont{Merlo}},
  \bibinfo{author}{\bibfnamefont{S.}~\bibnamefont{Pokorski}}, \bibnamefont{and}
  \bibinfo{author}{\bibfnamefont{M.}~\bibnamefont{Szleper}},
  \bibinfo{journal}{JHEP} \textbf{\bibinfo{volume}{07}}, \bibinfo{pages}{021}
  (\bibinfo{year}{2019}), \eprint{1905.03354}.

\bibitem[{\citenamefont{Lang et~al.}(2021)\citenamefont{Lang, Liebler,
  Sch\"afer-Siebert, and Zeppenfeld}}]{Lang:2021hnd}
\bibinfo{author}{\bibfnamefont{J.}~\bibnamefont{Lang}},
  \bibinfo{author}{\bibfnamefont{S.}~\bibnamefont{Liebler}},
  \bibinfo{author}{\bibfnamefont{H.}~\bibnamefont{Sch\"afer-Siebert}},
  \bibnamefont{and}
  \bibinfo{author}{\bibfnamefont{D.}~\bibnamefont{Zeppenfeld}},
  \bibinfo{journal}{Eur. Phys. J. C} \textbf{\bibinfo{volume}{81}},
  \bibinfo{pages}{659} (\bibinfo{year}{2021}), \eprint{2103.16517}.

\bibitem[{\citenamefont{Aaboud et~al.}(2017)}]{ATLAS:2016snd}
\bibinfo{author}{\bibfnamefont{M.}~\bibnamefont{Aaboud}} \bibnamefont{et~al.}
  (\bibinfo{collaboration}{ATLAS}), \bibinfo{journal}{Phys. Rev. D}
  \textbf{\bibinfo{volume}{96}}, \bibinfo{pages}{012007}
  (\bibinfo{year}{2017}), \eprint{1611.02428}.

\bibitem[{\citenamefont{Sirunyan et~al.}(2020{\natexlab{b}})}]{CMS:2020gfh}
\bibinfo{author}{\bibfnamefont{A.~M.} \bibnamefont{Sirunyan}}
  \bibnamefont{et~al.} (\bibinfo{collaboration}{CMS}), \bibinfo{journal}{Phys.
  Lett. B} \textbf{\bibinfo{volume}{809}}, \bibinfo{pages}{135710}
  (\bibinfo{year}{2020}{\natexlab{b}}), \eprint{2005.01173}.

\bibitem[{\citenamefont{Buchmueller et~al.}(2014)\citenamefont{Buchmueller,
  Dolan, and McCabe}}]{Buchmueller:2013dya}
\bibinfo{author}{\bibfnamefont{O.}~\bibnamefont{Buchmueller}},
  \bibinfo{author}{\bibfnamefont{M.~J.} \bibnamefont{Dolan}}, \bibnamefont{and}
  \bibinfo{author}{\bibfnamefont{C.}~\bibnamefont{McCabe}},
  \bibinfo{journal}{JHEP} \textbf{\bibinfo{volume}{01}}, \bibinfo{pages}{025}
  (\bibinfo{year}{2014}), \eprint{1308.6799}.

\bibitem[{\citenamefont{Henning et~al.}(2016)\citenamefont{Henning, Lu, and
  Murayama}}]{Henning:2014wua}
\bibinfo{author}{\bibfnamefont{B.}~\bibnamefont{Henning}},
  \bibinfo{author}{\bibfnamefont{X.}~\bibnamefont{Lu}}, \bibnamefont{and}
  \bibinfo{author}{\bibfnamefont{H.}~\bibnamefont{Murayama}},
  \bibinfo{journal}{JHEP} \textbf{\bibinfo{volume}{01}}, \bibinfo{pages}{023}
  (\bibinfo{year}{2016}), \eprint{1412.1837}.

\bibitem[{\citenamefont{Das~Bakshi
  et~al.}(2020{\natexlab{b}})\citenamefont{Das~Bakshi, Chakrabortty, and
  Spannowsky}}]{Bakshi:2020eyg}
\bibinfo{author}{\bibfnamefont{S.}~\bibnamefont{Das~Bakshi}},
  \bibinfo{author}{\bibfnamefont{J.}~\bibnamefont{Chakrabortty}},
  \bibnamefont{and}
  \bibinfo{author}{\bibfnamefont{M.}~\bibnamefont{Spannowsky}}
  (\bibinfo{year}{2020}{\natexlab{b}}), \eprint{2012.03839}.

\bibitem[{\citenamefont{Anisha et~al.}(2020)\citenamefont{Anisha, Das~Bakshi,
  Chakrabortty, and Patra}}]{Anisha:2020ggj}
\bibinfo{author}{\bibnamefont{Anisha}},
  \bibinfo{author}{\bibfnamefont{S.}~\bibnamefont{Das~Bakshi}},
  \bibinfo{author}{\bibfnamefont{J.}~\bibnamefont{Chakrabortty}},
  \bibnamefont{and} \bibinfo{author}{\bibfnamefont{S.~K.} \bibnamefont{Patra}}
  (\bibinfo{year}{2020}), \eprint{2010.04088}.

\bibitem[{\citenamefont{Das~Bakshi et~al.}(2019)\citenamefont{Das~Bakshi,
  Chakrabortty, and Patra}}]{Bakshi:2018ics}
\bibinfo{author}{\bibfnamefont{S.}~\bibnamefont{Das~Bakshi}},
  \bibinfo{author}{\bibfnamefont{J.}~\bibnamefont{Chakrabortty}},
  \bibnamefont{and} \bibinfo{author}{\bibfnamefont{S.~K.} \bibnamefont{Patra}},
  \bibinfo{journal}{Eur. Phys. J. C} \textbf{\bibinfo{volume}{79}},
  \bibinfo{pages}{21} (\bibinfo{year}{2019}), \eprint{1808.04403}.

\bibitem[{\citenamefont{Gherardi et~al.}(2020)\citenamefont{Gherardi, Marzocca,
  and Venturini}}]{Gherardi:2020det}
\bibinfo{author}{\bibfnamefont{V.}~\bibnamefont{Gherardi}},
  \bibinfo{author}{\bibfnamefont{D.}~\bibnamefont{Marzocca}}, \bibnamefont{and}
  \bibinfo{author}{\bibfnamefont{E.}~\bibnamefont{Venturini}},
  \bibinfo{journal}{JHEP} \textbf{\bibinfo{volume}{07}}, \bibinfo{pages}{225}
  (\bibinfo{year}{2020}), \bibinfo{note}{[Erratum: JHEP 01, 006 (2021)]},
  \eprint{2003.12525}.

\bibitem[{\citenamefont{Joglekar et~al.}(2012)\citenamefont{Joglekar,
  Schwaller, and Wagner}}]{Joglekar:2012vc}
\bibinfo{author}{\bibfnamefont{A.}~\bibnamefont{Joglekar}},
  \bibinfo{author}{\bibfnamefont{P.}~\bibnamefont{Schwaller}},
  \bibnamefont{and} \bibinfo{author}{\bibfnamefont{C.~E.~M.}
  \bibnamefont{Wagner}}, \bibinfo{journal}{JHEP} \textbf{\bibinfo{volume}{12}},
  \bibinfo{pages}{064} (\bibinfo{year}{2012}), \eprint{1207.4235}.

\bibitem[{\citenamefont{Angelescu and Huang}(2019)}]{Angelescu:2018dkk}
\bibinfo{author}{\bibfnamefont{A.}~\bibnamefont{Angelescu}} \bibnamefont{and}
  \bibinfo{author}{\bibfnamefont{P.}~\bibnamefont{Huang}},
  \bibinfo{journal}{Phys. Rev. D} \textbf{\bibinfo{volume}{99}},
  \bibinfo{pages}{055023} (\bibinfo{year}{2019}), \eprint{1812.08293}.

\bibitem[{\citenamefont{Chala et~al.}(2020)\citenamefont{Chala, Koz\'ow, Ramos,
  and Titov}}]{Chala:2020odv}
\bibinfo{author}{\bibfnamefont{M.}~\bibnamefont{Chala}},
  \bibinfo{author}{\bibfnamefont{P.}~\bibnamefont{Koz\'ow}},
  \bibinfo{author}{\bibfnamefont{M.}~\bibnamefont{Ramos}}, \bibnamefont{and}
  \bibinfo{author}{\bibfnamefont{A.}~\bibnamefont{Titov}},
  \bibinfo{journal}{Phys. Lett. B} \textbf{\bibinfo{volume}{809}},
  \bibinfo{pages}{135752} (\bibinfo{year}{2020}), \eprint{2005.09655}.

\bibitem[{\citenamefont{Bi\ss{}mann et~al.}(2021)\citenamefont{Bi\ss{}mann,
  Hiller, Hormigos-Feliu, and Litim}}]{Bissmann:2020lge}
\bibinfo{author}{\bibfnamefont{S.}~\bibnamefont{Bi\ss{}mann}},
  \bibinfo{author}{\bibfnamefont{G.}~\bibnamefont{Hiller}},
  \bibinfo{author}{\bibfnamefont{C.}~\bibnamefont{Hormigos-Feliu}},
  \bibnamefont{and} \bibinfo{author}{\bibfnamefont{D.~F.} \bibnamefont{Litim}},
  \bibinfo{journal}{Eur. Phys. J. C} \textbf{\bibinfo{volume}{81}},
  \bibinfo{pages}{101} (\bibinfo{year}{2021}), \eprint{2011.12964}.

\bibitem[{\citenamefont{Patel}(2017)}]{Patel:2016fam}
\bibinfo{author}{\bibfnamefont{H.~H.} \bibnamefont{Patel}},
  \bibinfo{journal}{Comput. Phys. Commun.} \textbf{\bibinfo{volume}{218}},
  \bibinfo{pages}{66} (\bibinfo{year}{2017}), \eprint{1612.00009}.

\end{thebibliography}

\end{document}